\pdfoutput=1
\documentclass[11pt]{article}
\usepackage[T1]{fontenc}
\usepackage[utf8]{inputenc}
\usepackage{ACL2023}
\usepackage{custom}

\usepackage{times}
\usepackage{microtype}
\usepackage{stmaryrd} 
\usepackage{latexsym}
\usepackage{inconsolata}
\usepackage[scaled=0.9]{helvet}
\usepackage{newtxmath}

\title{Convergence and Diversity in the Control Hierarchy}
\newcommand{\mailto}[2]{\texttt{\href{mailto:#1}{#2}}}

\author{Alexandra Butoi$^1$ \quad
Ryan Cotterell$^1$ \quad
David Chiang$^2$ \\[1ex]
  $^1$ETH Zürich \quad
  $^2$University of Notre Dame \\[1ex]
    \{\mailto{alexandra.butoil@inf.ethz.ch}{alexandra.butoi}, \mailto{ryan.cotterell@inf.ethz.ch}{ryan.cotterell}\}\texttt{@inf.ethz.ch} \quad
    \mailto{dchiang@nd.edu}{dchiang@nd.edu}
}

\begin{document}
\maketitle 

\begin{abstract}
\citeauthor{weir-1992-geometric} has defined a hierarchy of language classes whose second member (\tal) is generated by tree-adjoining grammars (TAG), linear indexed grammars (LIG), combinatory categorial grammars, and head grammars. The hierarchy is obtained using the mechanism of \emph{control}, and \tal{} is obtained using a context-free grammar (CFG) whose derivations are controlled by another CFG. We adapt Weir's definition of a controllable CFG to give a definition of controllable pushdown automata (PDAs). This yields three new characterizations of \tal{} as the class of languages generated by PDAs controlling PDAs, PDAs controlling CFGs, and CFGs controlling PDAs. We show that these four formalisms are not only weakly equivalent but equivalent in a stricter sense that we call \emph{\bijeqNN}. Furthermore, using an even stricter notion of equivalence called \emph{\topeqNN}, we make precise the intuition that a CFG controlling a CFG is a TAG, a PDA controlling a PDA is an embedded PDA, and a PDA controlling a CFG is a LIG. The fourth member of this family, a CFG controlling a PDA, does not correspond to any formalism we know of, so we invent one and call it a Pushdown Adjoining Automaton.
\end{abstract}
\section{Introduction}

\citet{weir-1992-geometric} defined a hierarchy of formal languages whose first level ($\mathcal{L}_1$) is the class of context-free languages, and whose second level (\tal{}) is the class generated by several existing formalisms: tree-adjoining grammars \cite{JOSHI1975136}, linear indexed grammars \cite{gazdar1988applicability}, combinatory categorial grammars \cite{steedman1987} and head grammars \cite{pollard1984}, which were proven weakly equivalent in a classic paper \citep{joshi-convergence-1991,sussex504}, and embedded pushdown automaton (EPDA) as well \citep{vijay-shanker1987}.

Weir's hierarchy is obtained using the mechanism of \emph{control}, and \tal{} is obtained by using a context-free grammar (CFG) to control another modified CFG, called a \emph{labeled distinguished CFG} or LD-CFG. Here, we define a controllable pushdown automaton (PDA), called a \emph{labeled distinguished PDA} (LD-PDA), and show that PDAs can also be used as controllers. By combining a controller CFG or PDA with a controlled LD-CFG or LD-PDA, we obtain a total of four formalisms that characterize~\tal{}.

\begin{figure}[t]
\centering
\begin{tikzpicture}[x=1.75cm,y=1cm]
\node(cfg-cfg) at (-0.2,0) {$\text{CFG} \control \text{CFG}$};
\node(cfg-pda) at (-0.2,1) {$\text{CFG} \control \text{PDA}$};
\node(pda-cfg) at (1.2,0) {$\text{PDA} \control \text{CFG}$};
\node(pda-pda) at (1.2,1) {$\text{PDA} \control \text{PDA}$};
\node(tag) at (-1,-1) {spinal TAG};
\node(lig) at (2,-1) {LIG};
\node(epda) at (2,2) {EPDA};
\node(paa) at (-1,2) {spinal PAA};
\begin{scope}
\draw (cfg-cfg) -- (cfg-pda);
\draw (cfg-pda) -- (pda-pda);
\draw (pda-pda) -- (pda-cfg);
\draw (pda-cfg) -- (cfg-cfg);
\end{scope}
\begin{scope}[very thick]
\draw (cfg-cfg) -- (tag);
\draw (cfg-pda) -- (paa);
\draw (pda-pda) -- (epda);
\draw (pda-cfg) -- (lig);
\end{scope}
\end{tikzpicture}
\caption{Overview of results. ``$X \control Y$'' means ``$X$ controlling $Y$.'' Thin lines indicate \bijeqNN{}, while thick lines indicate \topeqNN.}
\label{fig:overview}
\end{figure}

We show that these four formalisms are not only weakly equivalent but equivalent in a stricter sense that we call \bijeqNN{}. Furthermore, using an even stricter notion of equivalence called \topeqNN{}, we make precise the intuition that a CFG controlling a CFG is essentially TAG, a PDA controlling a PDA is an EPDA, and a PDA controlling a CFG is a LIG. The fourth member of this family, CFGs controlling PDAs, does not correspond to any existing automaton we know of, so we invent one and call it a Pushdown Adjoining Automaton (PAA).


The main contributions of this paper are:
\begin{compactitem}
    \item Adapting Weir's LD-CFG to pushdown automata (LD-PDA).
    \item Two new definitions of equivalence between formalisms, \bijeqNN{} and \topeqNN{}.
    \item Four \bijeqJJ{} formalisms (one old, three new) obtained by controlling an LD-CFG/PDA using a controller CFG/PDA, three of which are \topeqJJ{} to TAG, LIG and EPDA, respectively.
    \item A new formalism, PAA, \topeqJJ{} to the fourth formalism.
\end{compactitem}



\section{Preliminaries}

Let $[n]$ denote the set of integers $\{1, \ldots, n\}$, and let $\range{i}{j}$ denote the set of integers $\{i, \ldots, j\}$.


For any sets $\mathcal{X}$ and $\mathcal{Y}$, we write $ \dist{\mathcal{X}} = \{ \dist{\ntvar{X}} \mid \symvar{X} \in \mathcal{X}\}$ and $\mathcal{X}[\mathcal{Y}] = \{ \ntvar{X}[\ntvar{Y}] \mid \ntvar{X} \in \mathcal{X}, \ntvar{Y} \in \mathcal{Y} \}$.

\subsection{Context-free grammars}

We assume familiarity with CFGs; see \cref{sec:cfg} for definitions. The following normal form will be convenient later.

\begin{defin}
    A CFG is in \defn{\normalform{}} if its productions have one of the forms $\ntvar{X}\rightarrow \ntvar{Y}_1 \cdots \ntvar{Y}_k$ or $\ntvar{X}\rightarrow \symvar{a}$.
\end{defin}

A CFG derivation is sometimes thought of as a sequence of rewriting steps, and sometimes as a tree. The distinction is important in this paper, and we always refer to the former as \emph{derivations} and the latter as \emph{derivation trees}.
See \cref{sec:cfg}, \cref{def:cfg-derivation-tree} for a definition and \cref{fig:cfg-pda-d-strong-eq}a for an example.

\subsection{Pushdown automata}

\iffalse
\begin{defin}
A \defn{pushdown automaton} (PDA) is a tuple $\pushdown = (\states, \alphabet, \trans, (q_\init, \stackseq_{\init}), (q_\final, \stackseq_{\final}))$ where $\states$, $\alphabet$ and $\stackalphabet$ are finite sets of \emph{states}, \emph{input symbols} and \emph{stack symbols}, $\trans$ is a finite set of transitions, and $(q_\init, \stackseq_{\init})$ and $(q_\final, \stackseq_{\final}) \in \states \times \kleene\stackalphabet$ are the \emph{initial} and \emph{final configurations}. The transitions are of the form $p, \stackseq \xrightarrow{a} q, \stackseq'$, where $p, q \in \states$, $\stackseq, \stackseq' \in \kleene{\stackalphabet}$ and $\symvar{a} \in \alphabet \cup \{\varepsilon\}$. We call a transition \emph{scanning} iff $|a| > 0$, and \emph{$k$-pop, $l$-push} if $|\stackseq|=k$ and $|\stackseq'|=l$.
\end{defin}
\else 
\begin{defin}
A \defn{pushdown automaton} (PDA) is a tuple $\pushdown = (\states, \alphabet, \stackalphabet, \trans, (q_\init, \ntvar{S}), (q_\final, \varepsilon))$ where $\states$, $\alphabet$ and $\stackalphabet$ are finite sets of \emph{states}, \emph{input symbols} and \emph{stack symbols}, $\trans$ is a finite set of \emph{transitions}, and $(q_\init, \ntvar{S})$ and $(q_\final, \varepsilon)$ are called the \emph{initial} and \emph{final configurations}, where $q_{\init}, q_{\final}\in \states$ and $\ntvar{S}\in\stackalphabet$. The transitions are of the form $p, \ntvar{X} \xrightarrow{a} q, \stackseq$, where $p, q \in \states$, $\stackseq \in \kleene{\stackalphabet}$, $\ntvar{X}\in\stackalphabet$ and $\symvar{a} \in \alphabet \cup \{\varepsilon\}$. We say that a transition \emph{scans} $\symvar{a}$ and is \emph{scanning} iff $|\symvar{a}| > 0$.
\end{defin}
This definition is similar to that of \citet{hopcroft-2006-introduction} but has only a single final configuration. 
\fi
Stacks are represented as strings over $\stackalphabet$, from top to bottom. Thus, in the stack $\stackseq=\ntvar{X}_1 \cdots \ntvar{X}_n$, the $\ntvar{X}_1$ is at the top of the stack, while $\ntvar{X}_n$ is at the bottom.

See \cref{sec:pda} for further definitions. We again define a normal form that will be convenient later.
\iffalse
\begin{defin}
A PDA is in \defn{2-normal form} if its scanning transitions are $1$-pop, $0$-push and its non-scanning transitions are $1$-pop, $k$-push where $k \leq 2$. Moreover, the initial configuration is $(q_\init, \ntvar{S})$ and the final configuration is $(q_\final, \varepsilon)$ for some $q_\init, q_\final \in \states$ and $\ntvar{S}\in \stackalphabet$.
\end{defin}
\else
\begin{defin}
A PDA is in \defn{\normalform{}} if its transitions have one of the forms $q, \ntvar{X} \xrightarrow{\varepsilon} \stackseq$ or $q, \ntvar{X} \xrightarrow{\symvar{a}} \varepsilon$.
\end{defin}
\fi

Unlike with CFGs, there is essentially no difference between treating PDA derivations as sequences of transitions or as (unary-branching) trees. See \cref{sec:pda}, \cref{def:pda-derivation-tree} for a definition and \cref{fig:cfg-pda-d-strong-eq}b for an example.

\section{Equivalence of Formalisms}
\label{sec:equivalence}

We think of grammars and automata generically as ``formal systems'' that have derivation trees yielding strings.
By equipping formal systems with derivation trees, we can define a notion of equivalence that is stricter than weak equivalence, and another notion that is more precise than strong equivalence.

\newcommand{\relabel}{\lambda}
\newcommand{\reorder}{\pi}

\begin{defin}
    A \defn{formal system} $\system$ over alphabet~$\alphabet$ is a set $\carrierset$ of \defn{derivation trees} together with a \defn{yield function} $\yield \colon \carrierset \rightarrow \kleene\alphabet$
    that defines a language $\mathcal{L}(\mathbf{\system}) = \{\yield(\derivation) \mid \derivation \in \carrierset\}$.
\end{defin}

\begin{defin}
    Two formal systems $\system$ and $\system'$ are \defn{\bijeqJJ{}} if there is a bijection $\phi$ between the derivation trees of $\system$ and $\system'$ such that for any derivation tree $\derivation$ of $\system$ that yields the string $\str$, the derivation tree $\phi(\derivation)$ of $\system'$ also yields $\str$.
\end{defin}

\begin{example}
Consider two CFGs $\grammar_\textrm{fin}$ and $\grammar_\textrm{inf}$:
\begin{align*}
\grammar_\textrm{fin} &= \{ \nt{S} \rightarrow \sym{a} \} &
\grammar_\textrm{inf} &= \{ \nt{S} \rightarrow \nt{S}, \nt{S} \rightarrow \sym{a} \}.
\end{align*}
These grammars are weakly equivalent, but not \bijeqJJ, because $\grammar_\textrm{fin}$ has just one derivation of\/ $\sym{a}$, but $\grammar_\textrm{inf}$ has infinitely many, so there is no bijection between their derivation sets. 
\end{example}

Although there is a standard method to remove unary productions $\nt{S} \rightarrow \nt{S}$, on weighted grammars this method requires the semiring of weights to be complete; it will not work with general semirings. \BijeqNN{} captures this distinction.

\begin{example} \label{ex:d-weak-eq}
    Consider a CFG $\grammar_\sym{aa} = \{ \nt{S}\rightarrow\nt{A}\nt{A}; \nt{A}\rightarrow \sym{a} \}$ and PDA $\pushdown_\sym{aa} = \{ p,\nt{S}\xrightarrow{\varepsilon}p,\nt{AA}; p,\nt{A}\xrightarrow{\sym{a}}p,\varepsilon \}$. Both $\grammar_\sym{aa}$ and $\pushdown_\sym{aa}$ derive just one string, $\sym{aa}$, each with a single derivation. Thus, $\grammar_\sym{aa}$ and $\pushdown_\sym{aa}$ are \bijeqJJ.
\end{example}

Strong equivalence is traditionally understood to mean that two formal systems generate the same sets of structural descriptions, but this notion cannot be made precise without defining what structural descriptions are. For our purposes, it will suffice to use derivation trees, thought of as unlabeled, unordered trees.

\begin{defin}
    Two formal systems $\system$ and $\system'$ are \defn{\topeqJJ{}} if there is a bijection $\phi$ between the derivation trees of $\system$ and $\system'$ such that for any derivation tree $\derivation$ of $\system$,
    $\derivation$ and $\phi(\derivation)$ are isomorphic as unlabeled, unordered trees, and
    $\derivation$ and $\phi(\derivation)$ yield the same string.
\end{defin}

\begin{example}
\cref{fig:cfg-pda-d-strong-eq} shows the derivation trees of the string $\Sym{a} \Sym{a}$ in the CFG $\grammar$ and the PDA $\pushdown$ defined in \cref{ex:d-weak-eq}. $\grammar$ and $\pushdown$ are \emph{not} \topeqJJ{}, because there is no isomorphism between their derivation trees: the derivation tree in $\grammar$ is branching while the derivation tree in $\pushdown$ is not.
\begin{figure}
\centering\small\tikzset{level distance=22pt}
(a)
        \Tree [.{$\nt{S}$} [.{$\nt{A}$} {$\sym{a}$} ] [.{$\nt{A}$} {$\sym{a}$} ] ]
\hspace{3em}
(b)
        \Tree [.{$( p,\nt{S})$} \edge node[edge label] {$\varepsilon$}; [.{$( p,\nt{AA})$} \edge node[edge label] {$\sym{a}$}; [.{$( p,\nt{A})$} \edge node[edge label] {$\sym{a}$}; {$( p,\varepsilon)$} ] ] ]
    \caption{Example derivation trees for \cref{ex:d-weak-eq}. (a)~For CFG $\grammar$, the nodes are nonterminals/terminals. (b)~For PDA $\pushdown$, the nodes are configurations.}
    \label{fig:cfg-pda-d-strong-eq}
\end{figure} 
\end{example}

In general, CFG and PDA are \bijeqJJ{} (\cref{app:d-weak-eq}, \cref{proposition:cfg-pda-dweak-eq}), but not \topeqJJ.
\section{Control}

After a number of formalisms were proven to be equivalent characterizations of a language class \tal{} that is slightly more powerful than context-free languages (CFLs), \citet{weir-1992-geometric} introduced the control hierarchy to characterize an infinite sequence of language classes, starting with CFLs and \tal. The essential idea is that \tal{} languages can be recognized by a CFG in which the productions do not apply freely, but are \emph{controlled} by another CFG. 

\subsection{LD-CFGs}

To make a CFG controllable, \citeauthor{weir-1992-geometric} augments it slightly, as follows.

\begin{defin}
A \defn{labeled distinguished context-free grammar} (LD-CFG) is a tuple $\grammar = (\nonterm, \alphabet, \labelset, \rules, \ntvar{S})$, where $\nonterm$, $\alphabet$ and $\labelset$ are finite sets of \emph{nonterminal symbols}, \emph{terminal symbols} and \emph{labels}, $\rules$ is a finite set of \emph{productions}, and $\ntvar{S} \in \nonterm$ is the \emph{start symbol}.
The productions have one of the forms
$\ell:\ntvar{A} \rightarrow \alpha \dist{\ntvar{B}} \beta$ or 
$\ell:\ntvar{A} \rightarrow \alpha$, 
where $\ell\in\labelset$, $\ntvar{A} \in \nonterm$, $\dist{\ntvar{B}} \in \dist\nonterm$, and $\alpha, \beta \in \kleene{(\nonterm \cup \alphabet)}$.
\end{defin}

The labels enable a \emph{controller}, which is a formal system over $\labelset$, to decide what productions the LD-CFG (the \emph{controllee}) can use. Because the controller generates/accepts strings over $\labelset$, whereas an LD-CFG derivation tree has a branching structure, the distinguishing marks ($\dist{B}$) determine the paths of the derivation tree that are controlled.

\begin{defin}
A \defn{sentential form} of an LD-CFG is a pair $(\sent, \ctrllang)$ where $\sent \in \kleene{((\nonterm \cup \alphabet)[\kleene{\labelset}])}$ is a sequence of terminal and nonterminal symbols, each augmented with a string of labels, and $\ctrllang \subseteq \kleene{\labelset}$ is a set of \defn{control words}.
\end{defin}

\begin{defin}
If there is a production $\ell:\ntvar{A}\rightarrow \boldsymbol{\beta}_1\dist{\ntvar{B}}\boldsymbol{\beta}_2\in\rules$,
then we write
$(\sent_1\,\ntvar{A}[\ctrlword]\,\sent_2, \ctrllang) \xRightarrow{\ell} (\sent_1\,\boldsymbol{\beta}_1[\varepsilon]\,\ntvar{B}[\ctrlword\ell]\,\boldsymbol{\beta_2}[\varepsilon]\,\sent_2, \ctrllang)$.

If there is a production $\ell\colon\ntvar{A}\rightarrow\boldsymbol{\beta}$, where $\boldsymbol{\beta}\in \kleene{\nonterm}$,
then we write
$(\sent_1\,\ntvar{A}[\ctrlword]\,\sent_2, \ctrllang) \xRightarrow{\ell} (\sent_1\,\boldsymbol{\beta}[\varepsilon]\,\sent_2, \ctrllang\cup \{\ctrlword\ell \})$.
\end{defin}

\begin{defin}
Let $\ctrllang \subseteq \kleene\labelset$.
If \[(\ntvar{S}[\varepsilon], \emptyset) \xRightarrow{\ell_1} (\sent_1, \ctrllang_1) \xRightarrow{\ell_2} \cdots \Rightarrow{\ell_n} (\sent_n, \ctrllang_n)\]
where $\sent_n \in \kleene\alphabet$ and $\ctrllang_n \subseteq \ctrllang$, we say that $\grammar$ \defn{derives} $\sent_n$ under $\ctrllang$.
\end{defin}

\begin{example} \label{ex:controlled_cfg}
Below is an LD-CFG $\grammar_1$ and the derivation tree for the string $\sym{aabbccdd}$.
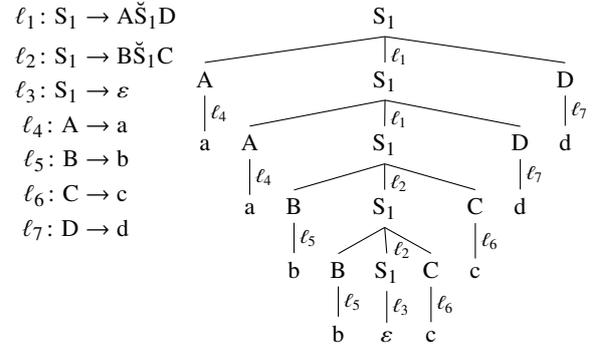
\[
\small
\begin{aligned}[t]
\Sym{\ell}_1 \colon \nt{S}_1 &\rightarrow \nt{A} \dist{\nt{S}}_1 \nt{D} \\
\Sym{\ell}_2 \colon \nt{S}_1 &\rightarrow \nt{B} \dist{\nt{S}}_1 \nt{C} \\
\Sym{\ell}_3 \colon \nt{S}_1 &\rightarrow \varepsilon \\
\Sym{\ell}_4 \colon \nt{A} &\rightarrow \sym{a} \\
\Sym{\ell}_5 \colon \nt{B} &\rightarrow \sym{b} \\
\Sym{\ell}_6 \colon \nt{C} &\rightarrow \sym{c} \\
\Sym{\ell}_7 \colon \nt{D} &\rightarrow \sym{d}
\end{aligned}%
\hspace{0.5em}%
\tikzset{sibling distance=0pt,level distance=24pt}
\tikzset{dist/.style={}} 
\Tree [.$\nt{S_1}$ 
        [.$\nt{A}$ \edge[dist] node[edge label] {$\Sym{\ell_4}$}; $\sym{a}$ ]
        \edge[dist] node[edge label,near end] {$\Sym{\ell_1}$}; 
        [.$\nt{S_1}$ 
          [.$\nt{A}$ \edge[dist] node[edge label] {$\Sym{\ell_4}$}; $\sym{a}$ ]
          \edge[dist] node[edge label,near end] {$\Sym{\ell_1}$}; 
          [.$\nt{S_1}$ 
            [.$\nt{B}$ \edge[dist] node[edge label] {$\Sym{\ell_5}$}; $\sym{b}$ ]
            \edge[dist] node[edge label,near end] {$\Sym{\ell_2}$}; 
            [.$\nt{S_1}$ 
              [.$\nt{B}$ \edge[dist] node[edge label] {$\Sym{\ell_5}$}; $\sym{b}$ ]
              \edge[dist] node[edge label,pos=1.5] {$\Sym{\ell_2}$};
              [.$\nt{S_1}$ \edge node[edge label] {$\Sym{\ell_3}$}; $\varepsilon$ ] 
              [.$\nt{C}$ \edge[dist] node[edge label] {$\Sym{\ell_6}$}; $\sym{c}$ ]
            ] 
            [.$\nt{C}$ \edge[dist] node[edge label] {$\Sym{\ell_6}$}; $\sym{c}$ ]
          ] 
          [.$\nt{D}$ \edge[dist] node[edge label] {$\Sym{\ell_7}$}; $\sym{d}$ ]
        ] 
        [.$\nt{D}$ \edge[dist] node[edge label] {$\Sym{\ell_7}$}; $\sym{d}$ ]
      ]
\]
The control words used in this derivation are $\{\Sym{\ell_1}\Sym{\ell_1}\Sym{\ell_2}\Sym{\ell_2}\Sym{\ell_3}, \Sym{\ell_4}, \Sym{\ell_5}, \Sym{\ell_6}, \Sym{\ell_7}\}$.
\end{example}

\begin{defin}
Let $\ctrllang \subseteq \kleene\labelset$. The \defn{language} of $\grammar$ under $\ctrllang$ is the set of strings $\mathcal{L}(\ctrllang,\grammar)=\{\str \mid \text{$\grammar$ derives $\str$ under $\ctrllang$}\}$.
\end{defin}

\begin{defin}
Let $\system_2$ be a formal system defining strings over $\labelset$ and $\grammar_1$ an LD-CFG. 
Then $\system_2$ controlling $\grammar_1$ form a single formal system, which we call $\system_2 \control \grammar_1$ and which defines the language
$\mathcal{L}(\system_2\control\grammar_1) = \mathcal{L}(\mathcal{L}(\system_2),\grammar_1)$. 
\end{defin}

\begin{example} Below is an example CFG $\grammar_2$ that can be used as a controller for $\grammar_1$ above (\cref{ex:controlled_cfg}).
\begin{align*}
\NT{S}_2 &\rightarrow \NT{T} \NT{L}_3 \\
\NT{T} &\rightarrow \NT{L}_1 \NT{T} \NT{L}_2 &
\NT{T} &\rightarrow \varepsilon \\
\NT{L}_i &\rightarrow \Sym{\ell}_i \quad (i \in \range{1}{3}) \\
\NT{S}_2 &\rightarrow \Sym{\ell}_i \quad (i \in \range{4}{7})
\end{align*}
It generates the language $\{\ell_1^n \ell_2^n \ell_3 \mid n \ge 0\} \cup \{\ell_i \mid i \in \range{4}{7} \}$, which makes the LD-CFG generate the language $\{\sym{a}^n\sym{b}^n\sym{c}^n\sym{d}^n \mid n \ge 0\}$.

The controller could also be a PDA, like $\pushdown_2$:
\begin{align*}
q, \NT{S}_2 &\xrightarrow{\varepsilon} q, \NT{T} \NT{L}_3 \\
q, \NT{T} &\xrightarrow{\varepsilon} q, \NT{L}_1 \NT{T} \NT{L}_2 &
q, \NT{T} &\xrightarrow{\varepsilon} q, \varepsilon \\
q, \NT{L}_i &\xrightarrow{\Sym{\ell}_i} q, \varepsilon\ (i \in \range{1}{3}) \\
q ,\NT{S}_2 &\xrightarrow{\Sym{\ell}_i} q, \varepsilon\ (i \in \range{4}{7})
\end{align*}
\end{example}


\subsection{LD-PDAs}

\citeauthor{weir-1992-geometric}'s definition of LD-CFG can be adapted to pushdown automata.
\begin{defin}
A \defn{labeled distinguished PDA} (LD-PDA) is a tuple $\pushdown = (\states, \alphabet, \stackalphabet, \labelset, \trans, (q_\init, \ntvar{S}), (q_\final, \stackseq_\final))$, where $\states$, $\alphabet$ and $\stackalphabet$ are finite sets of \emph{states}, \emph{input symbols} and \emph{stack symbols}, $\trans$ is a finite set of \emph{transitions}, and $(q_\init, \ntvar{S})$ and $(q_\final, \stackseq_\final) \in \states \times \kleene{(\stackalphabet[\kleene\labelset])}$ are the \emph{initial} and \emph{final configurations}.
The transitions are of the form 
$\Symvar{\ell}\colon q, \ntvar{A} \xrightarrow{\symvar{a}} r, \boldsymbol{\alpha}$,
or $\Symvar{\ell}\colon q, \ntvar{A} \xrightarrow{\symvar{a}} r, \boldsymbol{\alpha} \dist{\ntvar{B}} \boldsymbol{\beta}$, 
where $\Symvar{\ell} \in \labelset$, $q, r \in \states$, $\symvar{a} \in \alphabet$, $\ntvar{A} \in \stackalphabet$, $\dist{\ntvar{B}} \in \dist\stackalphabet$, and $\boldsymbol{\alpha}, \boldsymbol{\beta} \in \kleene\stackalphabet$.
\end{defin}

\begin{example}
Below is an example LD-PDA,~$\pushdown_1$, where $q$ is both initial and final:
\begin{align*}
\Sym{\ell}_1 \colon q, \nt{S}_1 &\xrightarrow{\varepsilon} q, \nt{A} \dist{\nt{S}}_1 \nt{D} &
\Sym{\ell}_2 \colon q, \nt{S}_1 &\xrightarrow{\varepsilon} q, \nt{B} \dist{\nt{S}}_1 \nt{C} \\
\Sym{\ell}_3 \colon q, \nt{S}_1 &\xrightarrow{\varepsilon} q, \varepsilon \\
\Sym{\ell}_4 \colon q, \nt{A} &\xrightarrow{\sym{a}} q, \varepsilon &
\Sym{\ell}_5 \colon q, \nt{B} &\xrightarrow{\sym{b}} q, \varepsilon \\
\Sym{\ell}_6 \colon q, \nt{C} &\xrightarrow{\sym{c}} q, \varepsilon &
\Sym{\ell}_7 \colon q, \nt{D} &\xrightarrow{\sym{d}} q, \varepsilon
\end{align*}
\end{example}

\begin{defin}
A \defn{configuration} of an LD-PDA $\pushdown$ is a tuple $(q,\stackseq,\ctrllang)$, where $q$ is the current state, $\stackseq \in \kleene{(\stackalphabet[\kleene\labelset])}$ is a sequence of stack symbols, each augmented with a string of labels, and $\ctrllang\subseteq\kleene{\labelset}$ is a set of control words.
\end{defin}

\begin{defin}
If there is a transition $\ell\colon q,\ntvar{A}\xrightarrow{\symvar{a}}r,\boldsymbol{\beta}_1\dist{\ntvar{B}}\boldsymbol{\beta}_2\in\trans$, 
we write $(q,\ntvar{A}[\ctrlword]\,\stackseq,\ctrllang)\xRightarrow{\ell}(r,\boldsymbol{\beta}[\varepsilon]\,\ntvar{B}[\ctrlword\ell]\,\boldsymbol{\beta}[\varepsilon]\,\stackseq',\ctrllang)$. 

If there is a transition $\ell\colon q, \ntvar{A}\xRightarrow{\symvar{a}} r, \boldsymbol{\beta}$, 
we write $(q,\ntvar{A}[\ctrlword]\,\stackseq,\ctrllang\cup \{\ctrlword\ell\})\xRightarrow{\ell}(r,\boldsymbol{\beta}[\varepsilon]\,\stackseq,\ctrllang\cup \{\ctrlword\ell\})$.
\end{defin}

\begin{defin}
Let $\ctrllang \subseteq \kleene\labelset$.
If $(q_\init, \ntvar{S}, \emptyset) \xRightarrow{\ell_1} (q_1, \stackseq_1, \ctrllang_1) \xRightarrow{\ell_2} \cdots \xRightarrow{\ell_{n-1}} (q_{n-1}, \stackseq_{n-1}, \ctrllang_{n-1}) \xRightarrow{\ell_n} (q_\final, \varepsilon, \ctrllang_n)$, where $\ctrllang_n \subseteq \ctrllang$, and for $i=1,\ldots,n$, transition $\ell_i$ scans $\symvar{a}_i$, then we say that $\pushdown$ \defn{accepts} string $\symvar{a}_1 \symvar{a}_2 \cdots \symvar{a}_{n-1} \symvar{a}_n$ under $\ctrllang$.
\end{defin}

\begin{defin}
Let $\ctrllang \subseteq \kleene\labelset$. The \defn{language} accepted by $\pushdown$ under $\ctrllang$ is the set of strings $\mathcal{L}(\ctrllang,\pushdown) = \{\str \mid \text{$\pushdown$ accepts $\str$ under $\ctrllang$}\}$.
\end{defin}

\begin{defin}
Let $\system_2$ be a formal system defining strings over $\labelset$ and $\pushdown_1$ an LD-PDA. 
Then $\system_2$ controlling $\pushdown_1$ form a single formal system, which we call $\system_2 \control \pushdown_1$ and which defines the language $\mathcal{L}(\system_2 \control \pushdown_1) = \mathcal{L}(\mathcal{L}(\system_2),\pushdown_1)$.
\end{defin}

\subsection{Four two-level formalisms}

Considering both CFGs and PDAs as both controllers and controllees yields four two-level formalisms, one of which is Weir's original two-level grammar, and the other three of which are new.

\begin{proposition} \label{thm:weak}
    CFG $\control$ CFG, CFG $\control$ PDA, PDA $\control$ PDA and PDA $\control$ CFG are \bijeqJJ{}.
\end{proposition}

\begin{proof}
    See \cref{app:d-weak-eq}.
\end{proof}

From now on, we will use different fonts for the symbols in the controllee and the controller. In a controllee LD-CFG/LD-PDA, we use $\ntvar{X}, \ntvar{Y}, \ldots$ and $\symvar{a}, \symvar{b}, \ldots$ for the nonterminal/stack symbols and terminal/input symbols. In a controller CFG/PDA, we use $\NTvar{A}, \NTvar{B}, \ldots$ for the nonterminal/stack symbols and $\Symvar{a}, \Symvar{b}, \ldots$ or $\Symvar{\ell}, \Symvar{\ell_1}, \ldots$ for the terminal/input symbols.

\section{Derivation Trees of Two-Level Formalisms} \label{sec:control-derivations}

In order to show that our four two-level formalisms are d-strongly equivalent to other \tal{} formalisms, we need to establish what derivation trees look like in these formalisms.
Weir actually gives two definitions of derivation in a two-level grammar. The first is the one we have reproduced above, where a partial control word is written inside the square brackets. The second runs a leftmost derivation of the controller CFG inside the square brackets.
In this section, we follow this second approach, except that we use derivation trees, not leftmost derivations. 
We give a separate definition for each two-level formalism.
Ideally, these definitions would have followed ``for free'' from the definitions of derivation trees of (LD-)CFGs and (LD-)PDAs, but we save this level of generality for future work.

\begin{figure*}[t]%
\centering
\scriptsize
\tikzset{every node/.style={inner xsep=0pt}}%
\tikzset{edge label/.style={auto,inner xsep=3pt}}%
\tikzset{sibling distance=0.5mm, level distance=8mm}%
\tikzset{edge from parent/.append style={color=black!80}}%
\begin{tikzpicture}
\node[anchor=north west] at (0cm,17.5cm) {\small CFG \control{} CFG};
\node[anchor=north] at (3.7cm,17.5cm) {%
\Tree [.{$\NTiii{S_2}{S_1}{}$}
  [.{$\NTiii{T}{S_1}{S_1}$}
    [.{$\NTiii{L_1}{S_1}{S_1}$}
      [.{$\NTiii{S_2}{A}{}$} $\sym{a}$ ]
      [.{$\NTiii{S_2}{D}{}$} $\sym{d}$ ]
    ]
  [.{$\NTiii{T}{S_1}{S_1}$}
    [.{$\NTiii{L_1}{S_1}{S_1}$}
      [.{$\NTiii{S_2}{A}{}$} $\sym{a}$ ]
      [.{$\NTiii{S_2}{D}{}$} $\sym{d}$ ]
    ]
    [.{$\NTiii{T}{S_1}{S_1}$} $\epsilon$ ]
    [.{$\NTiii{L_2}{S_1}{S_1}$}
      [.{$\NTiii{S_2}{B}{}$} $\sym{b}$ ]
      [.{$\NTiii{S_2}{C}{}$} $\sym{c}$ ]
    ]    
  ]
    [.{$\NTiii{L_2}{S_1}{S_1}$}
      [.{$\NTiii{S_2}{B}{}$} $\sym{b}$ ]
      [.{$\NTiii{S_2}{C}{}$} $\sym{c}$ ]
    ]    
  ]
  [.{$\NTiii{L_3}{S_1}{}$} $\epsilon$ ]
]%
};

\node[anchor=north west] at (7.8cm,17.5cm) {\small CFG \control{} PDA};
\node[anchor=north] at (11.9cm,17.5cm) {%
\Tree [.{$\NTiii{S_2}{S_1}{}$}
  [.{$\NTiii{T}{S_1}{S_1}$}
    [.{$\NTiii{L_1}{S_1}{S_1}$}
      [.{$\NTiii{S_2}{A}{}*\NTiii{S_2}{D}{}$}
        \edge node[edge label] {$\sym{a}$};
        [.{$\NTiii{S_2}{D}{}$} \edge node[edge label] {$\sym{d}$}; $\epsilon$ ]
      ]
    ]
  [.{$\NTiii{T}{S_1}{S_1}$}
    [.{$\NTiii{L_1}{S_1}{S_1}$}
      [.{$\NTiii{S_2}{A}{}*\NTiii{S_2}{D}{}$}
        \edge node[edge label] {$\sym{a}$};
        [.{$\NTiii{S_2}{D}{}$} \edge node[edge label] {$\sym{d}$}; $\epsilon$ ]
      ]
    ]
    [.{$\NTiii{T}{S_1}{S_1}$} $\epsilon$ ]
    [.{$\NTiii{L_2}{S_1}{S_1}$}
      [.{$\NTiii{S_2}{B}{}*\NTiii{S_2}{C}{}$}
        \edge node[edge label] {$\sym{b}$};
        [.{$\NTiii{S_2}{C}{}$} \edge node[edge label] {$\sym{c}$}; $\epsilon$ ]
      ]
    ]    
  ]
    [.{$\NTiii{L_2}{S_1}{S_1}$}
      [.{$\NTiii{S_2}{B}{}*\NTiii{S_2}{C}{}$}
        \edge node[edge label] {$\sym{b}$};
        [.{$\NTiii{S_2}{C}{}$} \edge node[edge label] {$\sym{c}$}; $\epsilon$ ]
      ]
    ]    
  ]
  [.{$\NTiii{L_3}{S_1}{}$} $\epsilon$ ]
]%
};

\node[anchor=north west] at (0cm,12cm) {\small PDA \control{} CFG};
\node[anchor=north] at (3.7cm,12cm) {%
\Tree [.{$\nt{S}_1[\NT{S}_2]$}
  [.{$\nt{S}_1[\NT{T} \NT{L}_3]$}
    [.{$\nt{S}_1[\NT{L}_1 \NT{T} \NT{L}_2 \NT{L}_3]$}
      [.{$\nt{A}[\NT{S}_2]$} $\sym{a}$ ]
      [.{$\nt{S}_1[\NT{T} \NT{L}_2 \NT{L}_3]$} 
        [.{$\nt{S}_1[\NT{L}_1 \NT{T} \NT{L}_2 \NT{L}_2 \NT{L}_3]$}
          [.{$\nt{A}[\NT{S}_2]$} $\sym{a}$ ]
          [.{$\nt{S}_1[\NT{T} \NT{L}_2 \NT{L}_2 \NT{L}_3]$} 
            [.{$\nt{S}_1[\NT{L}_2 \NT{L}_2 \NT{L}_3]$} 
              [.{$\nt{B}[\NT{S}_2]$} $\sym{b}$ ]
              [.{$\nt{S}_1[\NT{L}_2 \NT{L}_3]$}             
                [.{$\nt{B}[\NT{S}_2]$} $\sym{b}$ ]
                [.{$\nt{S}_1[\NT{L}_3]$} $\epsilon$ ]
                [.{$\nt{C}[\NT{S}_2]$} $\sym{c}$ ]
              ]
              [.{$\nt{C}[\NT{S}_2]$} $\sym{c}$ ]
            ]
          ]
          [.{$\nt{D}[\NT{S}_2]$} $\sym{d}$ ]
        ]
      ]
      [.{$\nt{D}[\NT{S}_2]$} $\sym{d}$ ]
    ]
  ]
]};

\node[anchor=north west] at (7.8cm,12cm) {\small PDA \control{} PDA};
\begin{scope}[xshift=9.6cm,yshift=11.1cm]
\Tree [.{$\nt{S}_1[\NT{S}_2]$}
[.{$\nt{S}_1[\NT{T} \NT{L}_3]$}
[.{$\nt{S}_1[\NT{L}_1 \NT{T} \NT{L}_2 \NT{L}_3]$}
[.{$\nt{A}[\NT{S}_2]\,\nt{S}_1[\NT{T} \NT{L}_2 \NT{L}_3]\,\nt{D}[\NT{S}_2]$}
\edge node[edge label] {$\sym{a}$};
[.{$\nt{S}_1[\NT{T} \NT{L}_2 \NT{L}_3]\,\nt{D}[\NT{S}_2]$}
[.{$\nt{S}_1[\NT{L}_1 \NT{T} \NT{L}_2 \NT{L}_2 \NT{L}_3]\,\nt{D}[\NT{S}_2]$}
[.{$\nt{A}[\NT{S}_2]\,\nt{S}_1[\NT{T} \NT{L}_2 \NT{L}_2 \NT{L}_3]\,\nt{D}[\NT{S}_2]\,\nt{D}[\NT{S}_2]$}
\edge node[edge label] {$\sym{a}$};
[.{$\nt{S}_1[\NT{T} \NT{L}_2 \NT{L}_2 \NT{L}_3]\,\nt{D}[\NT{S}_2]\,\nt{D}[\NT{S}_2]$}
[.\node(n1){$\nt{S}_1[\NT{L}_2 \NT{L}_2 \NT{L}_3]\,\nt{D}[\NT{S}_2]\,\nt{D}[\NT{S}_2]$};
] ] ] ] ] ] ] ] ]%
\end{scope}
\begin{scope}[xshift=13.8cm,yshift=11.1cm]
\Tree [.\node(n2){$\nt{B}[\NT{S}_2]\,\nt{S}_1[\NT{L}_2 \NT{L}_3]\,\nt{C}[\NT{S}_2]\,\nt{D}[\NT{S}_2]\,\nt{D}[\NT{S}_2]$};
\edge node[edge label] {$\sym{b}$};
[.{$\nt{S}_1[\NT{L}_2 \NT{L}_3]\,\nt{C}[\NT{S}_2]\,\nt{D}[\NT{S}_2]\,\nt{D}[\NT{S}_2]$}
[.{$\nt{B}[\NT{S}_2]\,\nt{S}_1[\NT{L}_3]\,\nt{C}[\NT{S}_2]\,\nt{C}[\NT{S}_2]\,\nt{D}[\NT{S}_2]\,\nt{D}[\NT{S}_2]$}
\edge node[edge label] {$\sym{b}$};
[.{$\nt{S}_1[\NT{L}_3]\,\nt{C}[\NT{S}_2]\,\nt{C}[\NT{S}_2]\,\nt{D}[\NT{S}_2]\,\nt{D}[\NT{S}_2]$}
[.{$\nt{C}[\NT{S}_2]\,\nt{C}[\NT{S}_2]\,\nt{D}[\NT{S}_2]\,\nt{D}[\NT{S}_2]$}
\edge node[edge label] {$\sym{c}$};
[.{$\nt{C}[\NT{S}_2]\,\nt{D}[\NT{S}_2]\,\nt{D}[\NT{S}_2]$}
\edge node[edge label] {$\sym{c}$};
[.{$\nt{D}[\NT{S}_2]\,\nt{D}[\NT{S}_2]$}
\edge node[edge label] {$\sym{d}$};
[.{$\nt{D}[\NT{S}_2]$}
\edge node[edge label] {$\sym{d}$};
$\epsilon$
] ] ] ] ] ] ] ]%
\end{scope}
\draw[edge from parent,rounded corners] (n1.south) |- +(1.9cm,-3mm) |- +(3cm,7.1cm) -| (n2.north);
\end{tikzpicture}
\caption{Derivation trees of $\sym{aabbccdd}$ under the four two-level formalisms considered in this paper. For pushdown automata, states are omitted.}
\label{fig:aabbccdd}
\end{figure*}

\Cref{fig:aabbccdd} shows the derivation of string $\sym{aabbccdd}$ under the four two-level formalisms considered in this paper. The rest of this section discusses each formalism in turn. We assume all controllers and controllees are in \normalform{}; see \cref{app:control-formalisms} for the general case.

\paragraph{PDA $\control$ PDA}

For PDA, derivations and derivation trees are the same thing, so derivation trees of PDA $\control$ PDA follow Weir's definition straightforwardly. 
There are three cases to consider.

(a) If a derivation node is labeled $q,\ntvar{X}[s, \NTvar{A}]\,\nestedstack$, 
and $s, \NTvar{A} \xrightarrow{\Symvar{\ell}} q_\final, \varepsilon$ is a controller transition,
and $\Symvar{\ell}$ is a controllee transition $q, \ntvar{X} \xrightarrow{\symvar{a}} r, \varepsilon$, 
then the node can have child
\begin{center}
\begin{tikzpicture}
\Tree [.$q,{\ntvar{X}[s,\NTvar{A}]\,\nestedstack}$
\edge node[edge label]{$\symvar{a}$}; ${r, \nestedstack}$ ]
\end{tikzpicture}
\end{center}

(b) If a derivation node is labeled $q,\ntvar{X}[s, \NTvar{A}\boldsymbol\beta]\,\nestedstack$, 
and $s, \NTvar{A} \xrightarrow{\Symvar{\ell}} t, \varepsilon$ is a controller transition,
and $\Symvar{\ell}$ is a controllee transition $q, \ntvar{X} \xrightarrow{\varepsilon} r, \ntvar{Y_1} \cdots \ntvar{Y_{d-1}}\,\dist{\ntvar{Z}}\,\ntvar{Y_{d+1}} \cdots \ntvar{Y}_k$, 
then
\begin{center}
\begin{tikzpicture} \scriptsize
\Tree [.$q,{\ntvar{X}[s,\NTvar{A}\boldsymbol\beta]\,\nestedstack}$  ${r,\ntvar{Y}_1[q_\init,\NTvar{S_2}] \cdots \ntvar{Y}_{d-1}[q_\init,\NTvar{S_2}]\,\ntvar{Z}[t,\boldsymbol\beta]\,\ntvar{Y}_{d+1}[q_\init,\NTvar{S_2}] \cdots \ntvar{Y}_k[q_\init,\NTvar{S_2}]\,\nestedstack}$ ]
\end{tikzpicture}
\end{center}

(c) If a derivation node is labeled $q,\ntvar{X}[s,\NTvar{A} \boldsymbol\beta]\,\nestedstack$, 
and $s,\NTvar{A} \xrightarrow{\varepsilon} t,\stackseq$ is a controller transition,
then
\begin{center}
\begin{tikzpicture}
\Tree [.$q,\ntvar{X}[s,\NTvar{A}\boldsymbol\beta]\,\nestedstack$
$q,\ntvar{X}[t,\stackseq\boldsymbol\beta]\,\nestedstack$ ]
\end{tikzpicture}
\end{center}

\paragraph{PDA $\control$ CFG} This formalism is not much more difficult than CFG $\control$ CFG.

(a) If a derivation node is labeled $\ntvar{X}[q,\NTvar{A}]$, 
and $q,\NTvar{A} \xrightarrow{\Symvar{\ell}} q_\final,\varepsilon$ is a controller transition,
and $\Symvar{\ell}$ is a controllee production $\ntvar{X} \rightarrow \symvar{a}$, then
\begin{center}
\begin{tikzpicture}[level distance=22pt]
\Tree [.$\ntvar{X}[q,\NTvar{A}]$
$\symvar{a}$ ] 
\end{tikzpicture}
\end{center}

(b) If a derivation node is labeled $\ntvar{X}[q,\NTvar{A} \boldsymbol\beta]$, 
and $q,\NTvar{A} \xrightarrow{\Symvar{\ell}} r,\varepsilon$ is a controller transition,
and $\Symvar{\ell}$ is a controllee production $\ntvar{X} \rightarrow \ntvar{Y_1} \cdots \ntvar{Y_{d-1}}\,\dist{\ntvar{Z}}\,\ntvar{Y_{d+1}} \cdots \ntvar{Y_k}$, 
then
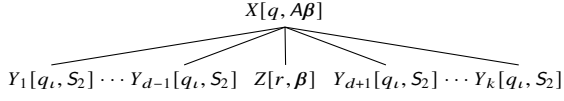
\begin{center} \scriptsize
\begin{tikzpicture}[sibling distance=6pt, every node/.append style={inner xsep=0pt}]
\Tree [.$\ntvar{X}[q,\NTvar{A}\boldsymbol\beta]$
$\ntvar{Y}_1[q_\init,\NTvar{S_2}]$ \edge[draw=none]; $\mathclap{\cdots}$ $\ntvar{Y}_{d-1}[q_\init,\NTvar{S_2}]$ $\ntvar{Z}[r,\boldsymbol\beta]$ $\ntvar{Y}_{d+1}[q_\init,\NTvar{S_2}]$ \edge[draw=none];  $\mathclap{\cdots}$ $\ntvar{Y}_k[q_\init,\NTvar{S_2}]$ ]
\end{tikzpicture}
\end{center}

(c) If a derivation node is labeled $\ntvar{X}[q,\NTvar{A} \boldsymbol\beta]$, 
and $q,\NTvar{A} \xrightarrow{\varepsilon} r,\stackseq$ is a controller transition,
then
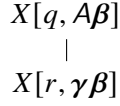
\begin{center}
\begin{tikzpicture}
\Tree [.$\ntvar{X}[q,\NTvar{A}\boldsymbol\beta]$
$\ntvar{X}[r,\stackseq\boldsymbol\beta]$ ]
\end{tikzpicture}
\end{center}

\paragraph{CFG $\control$ CFG}

\newcommand{\plug}[2]{#1(#2)}

When the controller is a CFG and therefore has branching derivation trees, the controllee's derivation becomes fragmented. To ensure that the fragments form one correct derivation, we need to add some extra information to the derivation nodes: each node is labeled $\NTvariii{A}{X}{Z}$, where $\NTvar{A}$ is a controller nonterminal, and $\ntvar{X}$ and $\ntvar{Z}$ are controllee nonterminals or $\bot$. We write $\NTvariii{A}{X}{\bot}$ as $\NTvariii{A}{X}{}$.

Moreover, the yield of a subderivation is in general a discontinuous string. We write $*$ to indicate a hole in a string (meant to resemble foot nodes in TAG), and we write $\plug{u}{v}$ to plug $v$ into the first hole in $u$; that is, if $u = u_1 * u_2$ where $u_1$ does not have a hole, then $\plug{u}{v} = u_1 v u_2$.

(a) If a derivation node is labeled $\NTvariii{A}{X}{}$,
and $\NTvar{A} \rightarrow \Symvar{\ell}$ is a controller production,
and $\Symvar{\ell}$ is a controllee production $\ntvar{X} \rightarrow \symvar{a}$, then
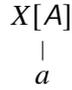
\begin{center}
\begin{tikzpicture}[level distance=22pt]
\Tree [.$\NTvariii{A}{X}{}$
$\symvar{a}$ ] 
\end{tikzpicture}
\end{center}
The yield is $a$.

(b) If a derivation node is labeled $\NTvariii{A}{X}{Z}$, 
and $\NTvar{A} \rightarrow \Symvar{\ell}$ is a controller production,
and $\Symvar{\ell}$ is a controllee production $\ntvar{X} \rightarrow \ntvar{Y_1} \cdots \ntvar{Y_{d_{i-1}}}\,\dist{\ntvar{Z}}\,\ntvar{Y_{d+1}} \cdots \ntvar{Y_k}$, 
then
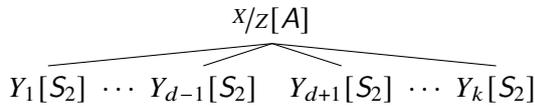
\begin{center}
\begin{tikzpicture}
\Tree [.$\NTvariii{A}{X}{Z}$ $\NTvariii{S_2}{Y_1}{}$ \edge[draw=none]; $\mathclap{\cdots}$ $\NTvariii{S_2}{Y_{d-1}}{}$  $\NTvariii{S_2}{Y_{d+1}}{}$ \edge[draw=none];  $\mathclap{\cdots}$ $\NTvariii{S_2}{Y_k}{}$  ]
\end{tikzpicture}
\end{center}
If each child node $\NTvariii{S_2}{Y_i}{}$ has yield $w_i$, then the parent has yield $w_1 \cdots w_{d-1} * w_{d+1} \cdots w_k$.

(c) If a derivation node is labeled $\NTvariii{A}{X}{Z}$,
and $\NTvar{A} \rightarrow \NTvar{B_1} \cdots \NTvar{B_k}$ is a controller production,
then
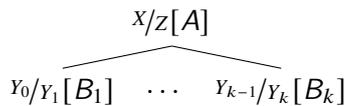
\begin{center}
\begin{tikzpicture}
\Tree [.$\NTvariii{A}{X}{Z}$
$\NTvariii{B_1}{Y_0}{Y_1}$ \edge[draw=none]; $\cdots$ $\NTvariii{B_k}{Y_{k-1}}{Y_k}$ ]
\end{tikzpicture}
\end{center}
where $\ntvar{Y_0} = \ntvar{X}$, $\ntvar{Y_k} = \ntvar{Z}$, and $\ntvar{Y_1}, \ldots, \ntvar{Y_{k-1}}$ can be any controllee nonterminals.
If the children have yields $w_1, w_2, \ldots, w_k$, then the parent has yield $\plug{w_1}{\plug{w_2}{\plug{\cdots}{w_k}}}$.

\paragraph{CFG $\control$ PDA}

The derivation trees of CFG $\control$ PDA are the most complicated. To cut down on complexity, we make use of the fact that every PDA that accepts by empty stack is equivalent to one with only one state \citep{hopcroft-2006-introduction}, and indeed the equivalence is d-strong (\cref{app:context-free-fs}, \cref{thm:onestatepda}). Therefore, we assume that the controllee has just one state, $q$.

Derivation nodes are labeled $(q, \nestedstack)$, where $q$ is a state and $\nestedstack$ is a string of symbols of the form $\NTvariii{A}{X}{Z}$ (as above, for CFG $\control$ CFG) or $*$.

\newcommand{\maybestar}{*^?}

(a) If a derivation node is labeled $q,\maybestar\NTvariii{A}{X}{}\,\nestedstack$, 
where $\maybestar$ denotes either the presence or absence of a $*$,
and $\NTvar{A} \rightarrow \Symvar \ell$ is a controller production,
and $\Symvar{\ell}$ is a controllee transition $q, \ntvar{X} \xrightarrow{\symvar{a}} q, \varepsilon$, 
then
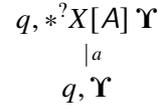
\begin{center}
\begin{tikzpicture}
\Tree [.${q,\maybestar\NTvariii{A}{X}{}\,\nestedstack}$
\edge node[edge label]{$\symvar{a}$}; ${q, \nestedstack}$ ]
\end{tikzpicture}
\end{center}
If the yield of the child is $w$, then the yield of the parent is $\maybestar aw$.

(b) If a derivation node is labeled $q,\NTvariii{A}{X}{Z}\,\nestedstack$,
and $\NTvar{A} \rightarrow \Symvar{\ell}$ is a controller production,
and $\Symvar{\ell}$ is a controllee transition $q, \ntvar{X} \xrightarrow{\varepsilon} q, \ntvar{Y_1} \cdots \ntvar{Y_{d-1}}\,\dist{\ntvar{Z}}\,\ntvar{Y_{d+1}} \cdots \ntvar{Y_k}$, then
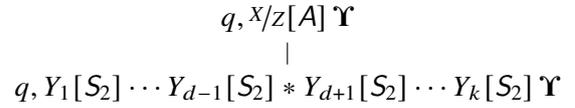
\begin{center}
\begin{tikzpicture}
\Tree [.${q,\NTvariii{A}{X}{Z}\,\nestedstack}$ ${q,\NTvariii{S_2}{Y_1}{} \cdots \NTvariii{S_2}{Y_{d-1}}{} * \NTvariii{S_2}{Y_{d+1}}{} \cdots \NTvariii{S_2}{Y_k}{}\,\nestedstack}$ ]
\end{tikzpicture}
\end{center}
The yield of the parent is just the yield of the child.

(c) If a derivation node is labeled $q,\NTvariii{A}{X}{Z}\,\nestedstack$,
and $\NTvar{A} \rightarrow \NTvar{B}_1 \cdots \NTvar{B}_k$ is a controller production,
then
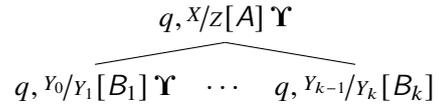
\begin{center}
\begin{tikzpicture}
\Tree [.$q,\NTvariii{A}{X}{Z}\,\nestedstack$
$q,\NTvariii{B_1}{Y_0}{Y_1}\,\nestedstack$ \edge[draw=none]; $\cdots$ $q,\NTvariii{B_k}{Y_{k-1}}{Y_k}$ ]
\end{tikzpicture}
\end{center}
where $\ntvar{Y_0} = \ntvar{X}$, $\ntvar{Y_k} = \ntvar{Z}$, and $\ntvar{Y_1}, \ldots, \ntvar{Y_{k-1}}$ can be any controllee nonterminals.
If the children have yields $w_1, w_2, \ldots, w_k$, then the parent has yield $\plug{w_1}{\plug{w_2}{\plug{\cdots}{w_k}}}$.

The reader may be surprised that $\nestedstack$ is placed on the leftmost child. 
It would be equally sensible to place it on any of the children, but the leftmost child makes the yield function simplest.

\section{Correspondences with \tal{} Formalisms}

We can now show how PDA \control{} PDA, PDA \control{} CFG, and CFG \control{} CFG are d-strongly equivalent to EPDA, LIG, and TAG, respectively, and to introduce PAA, which is d-strongly equivalent to CFG \control{} PDA.

\subsection{Embedded pushdown automata}

\iffalse
\begin{defin}\label{def:epda}
An \defn{embedded pushdown automaton}\footnote{Our definition slightly generalizes the standard definition of EPDA by allowing labeled inner stacks.} (EPDA) is a tuple $\pushdown = (\states, \alphabet, \nonterm, \stackalphabet, \trans, (q_\init, \nestedstack_\init),(q_\final, \nestedstack_\final))$, where $\states$, $\alphabet$, $\nonterm$ and $\stackalphabet$ are finite sets of \emph{states}, \emph{input symbols}, \emph{stack labels} and \emph{stack symbols}, $\trans$ is a finite set of \emph{transitions}, $(q_\init, \nestedstack_\init)$ and $(q_\final, \nestedstack_\final) \in \states \times \kleene{(\nonterm [\kleene{\stackalphabet}])}$ are called the \emph{initial} and \emph{final configurations}. The transitions are of the form $p,X[\stackseq]\xrightarrow{\symvar{a}} q,\epdastack{\nestedstack}{Y[\stackseq']}{\nestedstack'}$, where $p,q\in\states$, $X[\stackseq],Y[\stackseq']\in\nonterm[\kleene{\stackalphabet}]$, $\nestedstack,\nestedstack'\in \kleene{(\nonterm [\kleene{\stackalphabet}])}$, and $a \in \alphabet$.
\end{defin}
\david{I thought the transition form should be $p, X[\rest x] \xrightarrow{a} q, \nestedstack_1 X[\rest \stackseq] \nestedstack_2$.}
\else
\begin{defin}\label{def:epda}
An \defn{embedded pushdown automaton} (EPDA) is a tuple $\pushdown = (\states, \alphabet, \stackalphabet, \trans, (q_\init, [\NTvar{S}]),(q_\final, \varepsilon))$, where $\states$, $\alphabet$, $\nonterm$ and $\stackalphabet$ are finite sets of \emph{states}, \emph{input symbols} and \emph{stack symbols}, $\trans$ is a finite set of \emph{transitions}, and $(q_\init,[\NTvar{S}])$ and $(q_\final, \varepsilon)$ are called the \emph{initial} and \emph{final configurations}, where $q_\init, q _\final \in \states$, and $\NTvar{S}\in \stackalphabet$. The transitions are of the form $p, [\NTvar{A}\rest] \xrightarrow{\symvar{a}} q, \nestedstack [\stackseq \rest] \nestedstack'$ or $p, [\NTvar{A}] \xrightarrow{\symvar{a}} q, \varepsilon$, where $p,q\in\states$, $\NTvar{A}\in\stackalphabet$, $\stackseq \in \kleene{\stackalphabet}$, $\nestedstack,\nestedstack'\in \kleene{([\NTvar{S}])}$, and $\symvar{a} \in \alphabet \cup \{\varepsilon \}$.
\end{defin}
An EPDA maintains a stack of stacks. Each transition can pop and push from the top stack, as well as pop the top stack and/or push other stacks above and below it.
\fi

\begin{proposition} \label{thm:strong-epda}
    An EPDA is \topeqJJ{} to a PDA $\control$ PDA.
\end{proposition}

\begin{proof}
See \cref{app:d-strong-eq}.
\end{proof}

\subsection{Linear indexed grammars}

\iffalse
\begin{defin} \label{def:lig}
A \defn{linear indexed grammar}\footnote{Note that LIGs are not the same as the formalism of \citet{duske+parchmann:1984} by the same name.} (LIG) is a tuple $\grammar = (\nonterm, \alphabet, \stackalphabet, \rules, \ntvar{S}, \stackseq_{\init})$, where $\nonterm$, $\alphabet$ and $\stackalphabet$ are sets of \emph{nonterminals}, \emph{terminals} and \emph{stack symbols}, $\rules$ is a finite set of \emph{production rules}, $\ntvar{S} \in \nonterm$ is a distinguished \emph{start symbol}, and $\stackseq_{\init} \in \kleene\stackalphabet$ is the \emph{initial stack}. The production rules are of the form $\ntvar{A} [\rest \stackseq] \rightarrow \nestedstack\,\ntvar{A'}  [\rest \stackseq']\,\nestedstack'$ or $\ntvar{A} [\stackseq] \rightarrow \str$, where $\ntvar{A}, \ntvar{A'} \in \nonterm$, $\stackseq, \stackseq' \in \kleene{\stackalphabet}$, $\nestedstack, \nestedstack' \in \kleene{(\nonterm [\kleene{\stackalphabet}]\cup \alphabet)}$, and $\str \in \kleene{\alphabet}$. 
\end{defin}
\david{There seems to be a lot of variation in definitions. I'm surprised that this one doesn't allow rules $\ntvar{A}[\stackseq]\rightarrow\nestedstack$.}

This definition is adapted from \citet{sussex504} and is more general than other definitions.
Note that LIGs are not the same as the formalism of \citet{duske+parchmann:1984} by the same name.

\else
\begin{defin} \label{def:lig}
A \defn{linear indexed grammar} (LIG) is a tuple $\grammar = (\nonterm, \alphabet, \stackalphabet, \rules, \ntvar{S}, \NTvar{S})$, where $\nonterm$, $\alphabet$ and $\stackalphabet$ are sets of \emph{nonterminals}, \emph{terminals} and \emph{stack symbols}, $\rules$ is a finite set of \emph{productions}, $\ntvar{S} \in \nonterm$ is the \emph{start symbol}, and $\NTvar{S} \in \stackalphabet$ is the \emph{initial stack}. The production rules are of the form $\ligprod{X}{A}{\nestedstack_1}{Y}{\stackseq}{\nestedstack_2}$ or $\ligprodterm{X}{A}{a}$,
where $\ntvar{X}, \ntvar{Y} \in \nonterm$, $\stackseq \in \kleene{\stackalphabet}$, $\NTvar{A}\in\stackalphabet$, $\nestedstack_1, \nestedstack_2 \in \kleene{(\nonterm [\NTvar{S}]\cup \alphabet)}$, and $\symvar{a} \in \alphabet \cup \{\varepsilon \}$. 
\end{defin}
\fi

Each nonterminal symbol is augmented with a stack (in square brackets). A production with lhs $\ntvar{X}$ rewrites a nonterminal $\ntvar{X[\NTvar{A}\stackseq]}$ into its rhs, 
while popping $\NTvar{A}$ and pushing zero or more stack symbols.\footnote{Most LIG definitions \cite{vijay-shanker1987, joshi-convergence-1991, makoto2014} allow only productions that pop/push at most one symbol from/to a nonterminal's stack. } Exactly one of the rhs nonterminals inherits the stack, while the others get new stacks.

\begin{proposition} \label{thm:strong-lig}
    A LIG is \topeqJJ{} to a PDA $\control$ CFG.
\end{proposition}

\begin{proof}
See \cref{app:d-strong-eq}.
\end{proof}

\subsection{Tree-adjoining grammars}

\begin{figure*} \small
\tikzset{sibling distance=0pt, level distance=20pt}
\centering
\begin{tabular}{@{}ccc@{\hspace*{1cm}}l@{}}
\multicolumn{3}{@{}l}{TAG productions} & TAG derivation tree for $\sym{aabbccdd}$
\\[2ex]
$\begin{aligned}[t]
\beta_1 \colon \NT{S} &\rightarrow \Tree [.$\NT{T}$ $\NT{L_3}\foot$ ] \\
\beta_2 \colon \NT{T} &\rightarrow \Tree [.$\NT{L_1}$ [.$\NT{T}$ $\NT{L_2}\foot$ ] ] \\
\beta_3 \colon \NT{T} &\rightarrow \nt{X}\foot
\end{aligned}$
&
$\begin{aligned}[t]
\beta_4 \colon \NT{L_1} &\rightarrow \Tree [.$\nt{X}$ $\NT{A}\subst$ $\nt{X}\foot$ $\NT{D}\subst$ ] \\
\beta_5 \colon \NT{L_2} &\rightarrow \Tree [.$\nt{X}$ $\NT{B}\subst$ $\nt{X}\foot$ $\NT{C}$ ] \\
\beta_6 \colon \NT{L_3} &\rightarrow \Tree [.$\nt{X}$ $\epsilon$ ]
\end{aligned}$
&
$\begin{aligned}[t]
\beta_7 \colon \NT{A} &\rightarrow \Tree [.$\nt{X}$ $\sym{a}$ ] \\
\beta_8 \colon \NT{B} &\rightarrow \Tree [.$\nt{X}$ $\sym{b}$ ] \\
\beta_9 \colon \NT{C} &\rightarrow \Tree [.$\nt{X}$ $\sym{c}$ ] \\
\beta_{10} \colon \NT{D} &\rightarrow \Tree [.$\nt{X}$ $\sym{d}$ ]
\end{aligned}$
&
{\tikzset{level distance=30pt}
\Tree [.$\beta_1$ [.$\beta_2$ [.$\beta_4$ $\beta_7$ $\beta_{10}$ ] [.$\beta_2$ [.$\beta_4$ $\beta_7$ $\beta_{10}$ ] $\beta_3$ [.$\beta_5$ $\beta_8$ $\beta_9$ ] ] [.$\beta_5$ $\beta_8$ $\beta_9$ ] ] $\beta_6$ ]}
\end{tabular}

\caption{Example TAG and derivation tree for $\sym{aabbccdd}$. To reduce clutter, symbols have been renamed. Symbol~$\nt{X}$ is constant; other uppercase symbols are variable.}
\label{fig:tag_example}
\end{figure*}

\begin{defin}\label{def:tag-lang}
A \defn{tree-adjoining grammar} (TAG) is a tuple $\grammar = (\mathcal{V}, \mathcal{C}, \alphabet, \NTvar{S}, \rules)$ where $\mathcal{V}$, $\mathcal{C}$ and $\alphabet$ are finite sets of \emph{variable}, \emph{constant},\footnote{The terms \emph{variable} and \emph{constant} come from \citet{lang-1994-recognition} and are equivalent to nonterminals with obligatory and null adjunction constraints, respectively. 
}
and \emph{terminal symbols}, $\NTvar{S} \in \mathcal{V}$ is the \emph{start symbol} and $\rules$ is a set of \emph{productions}. The productions are of the form $\NTvar{X} \rightarrow \beta$, where $\NTvar{X}\in\mathcal{V}$ and $\beta \in \mathcal{T}(\mathcal{V} \cup \mathcal{C} \cup \alphabet \cup \{\varepsilon\})$. In~$\beta$, interior nodes must be in $\mathcal{V} \cup \mathcal{C}$. There is one leaf in $\mathcal{V} \cup \mathcal{C}$ that is designated the \defn{foot node} and marked with \raisebox{-3pt}{$\foot$}. The path from the root to the foot node is called the \defn{spine}. Other leaves must be in $\mathcal{V}$ (in which case they are called \defn{substitution nodes} and marked with \raisebox{2pt}{$\subst$}) or $\alphabet \cup \{\varepsilon\}$.
\end{defin}

This definition is close to \emph{non-strict TAG} \citep{lang-1994-recognition,rogers:2003}. Unlike the usual definition, its rules have left-hand sides and do not require the root and foot node to have the same label.
We also do not allow substitution, as it can be simulated by adjunction. A TAG derivation starts with 
$\NTvar{S}$
and proceeds by repeatedly choosing a variable $\NTvar{X}$ and rewriting it using a production $\NTvar{X} \rightarrow \beta$, such that the children of $\NTvar{X}$ become the children of $\beta$'s foot node. See \cref{app:tag} for additional definitions.

\begin{defin}
    A TAG production $\NTvar{X} \rightarrow \beta$, where $\beta$ has a foot node, is called \defn{spinal}\footnote{This term comes from \citet{fujiyoshi2000}, who use it for context-free tree grammars.} if every node of is either on the spine or the child of a node on the spine.
    A TAG production $\NTvar{X} \rightarrow \alpha$, where $\alpha$ does not have a foot node, is called spinal if there is a leaf $\eta$ such that every node is either on the path from the root to $\eta$ or the child of such a node. 
    A TAG is called spinal if all its productions are spinal.
\end{defin}


TAG derivation trees were defined by \citet{vijay-shanker1987}.
We show an example TAG and derivation tree in \cref{fig:tag_example}.

\begin{proposition} \label{thm:strong-tag}
    A spinal TAG is \topeqJJ{} to a CFG $\control$ CFG.
\end{proposition}

\begin{proof}
See \cref{app:d-strong-eq}; we give a brief sketch here, using the same three cases as in \cref{sec:control-derivations}.
Case (a) corresponds to the TAG production
\[
\NTvariii{A}{\ntvar{X}}{} \rightarrow
\begin{tikzpicture}[baseline=0,level distance=24pt]
\Tree [.$\ntvar{X}$
$\symvar{a}$ 
] 
\end{tikzpicture}
\]

Case (b) corresponds to the TAG production

{\small
\[ 
\NTvariii{A}{X}{Z} \rightarrow \hspace{-3em}
\begin{tikzpicture}[baseline=0,sibling distance=0pt]
\Tree [.$\ntvar{X}$ $\NTvariii{S_2}{Y_1}{}\subst$ \edge[draw=none]; $\mathclap{\cdots}$ $\NTvariii{S_2}{Y_{d-1}}{}\subst$ $\ntvar{Z}\foot$ $\NTvariii{S_2}{Y_{d+1}}{}\subst$ \edge[draw=none]; $\mathclap{\cdots}$ $\NTvariii{S_2}{Y_k}{}\subst$ ]
\end{tikzpicture}
\]}

Case (c) corresponds to the TAG productions
\begin{align*}
\NTvariii{A}{X}{Z} \rightarrow
\begin{tikzpicture}[baseline=0,level distance=24pt]
\Tree [.$\NTvariii{B_1}{X}{Y_1}$ [.$\NTvariii{B_2}{Y_1}{Y_2}$ \edge[dotted,very thick]; [.$\NTvariii{B_k}{Y_{k-1}}{Z}\foot$ ] ] ]
\end{tikzpicture}
&
\quad\NTvariii{A}{X}{} \rightarrow
\begin{tikzpicture}[baseline=0,level distance=24pt]
\Tree [.$\NTvariii{B_1}{X}{Y_1}$ [.$\NTvariii{B_2}{Y_1}{Y_2}$ \edge[dotted,very thick]; [.$\NTvariii{B_k}{Y_{k-1}}{}\subst$ ] ] ]
\end{tikzpicture}
\end{align*}
If $k=0$ then the corresponding TAG productions are
\begin{align*}
\NTvariii{A}{X}{X} &\rightarrow \ntvar{X}\foot &
\NTvariii{A}{X}{} &\rightarrow \Tree [.$\ntvar{X}$ $\varepsilon$ ]
\tag*{\qedhere}
\end{align*}

\end{proof}

\subsection{Pushdown adjoining automata}
\label{sec:paa}

\begin{figure*}[t] \small
\tikzset{paa picture}
\centering
\begin{tabular}{@{}ll@{}}
PAA transitions & PAA run for $\sym{aabbccdd}$ \\
$\begin{aligned}[t]
\begin{tikzpicture}[baseline=0]
\node(n1){$\NT{S}$};
\end{tikzpicture}
&\xrightarrow{\epsilon}
\begin{tikzpicture}[baseline=0]
\node(n1){$\NT{T}$};
\dnode{n1}{n2}{$\NT{L}_3$}
\end{tikzpicture}
&\begin{tikzpicture}[baseline=0]
\node at (0,0) (n1){$\NT{L}_3$};
\end{tikzpicture}
&\xrightarrow{\epsilon}
\epsilon
\\
\begin{tikzpicture}[baseline=0]
\node(n1){$\NT{T}$};
\fnode{n1}
\end{tikzpicture}
&\xrightarrow{\epsilon}
\begin{tikzpicture}[baseline=0]
\node(n1){$\NT{L}_1$};
\dnode{n1}{n2}{$\NT{T}$}
\dnode{n2}{n3}{$\NT{L}_2$}
\fnode{n3}
\end{tikzpicture}
&
\begin{tikzpicture}[baseline=0]
\node(n1){$\NT{T}$};
\fnode{n1}
\end{tikzpicture}
&\xrightarrow{\epsilon}
\begin{tikzpicture}[baseline=0]
\node(n1){$\nt{X}$};
\fnode{n1}
\end{tikzpicture}
\\
\begin{tikzpicture}[baseline=0]
\node(n1){$\NT{L}_1$};
\fnode{n1}
\end{tikzpicture}
&\xrightarrow{\epsilon}
\begin{tikzpicture}[baseline=0]
\node(n1){$\NT{A}$};
\rnode{n1}{n2}{$\nt{X}$}
\rnode{n2}{n3}{$\NT{D}$}
\fnode{n2}
\end{tikzpicture}
&
\begin{tikzpicture}[baseline=0]
\node(n1){$\NT{L}_2$};
\fnode{n1}
\end{tikzpicture}
&\xrightarrow{\epsilon}
\begin{tikzpicture}[baseline=0]
\node(n1){$\NT{B}$};
\rnode{n1}{n2}{$\nt{X}$}
\rnode{n2}{n3}{$\NT{C}$}
\fnode{n2}
\end{tikzpicture}
\\
\begin{tikzpicture}[baseline=0]
\node at (0,0) (n1){$\NT{A}$};
\end{tikzpicture}
&\xrightarrow{\sym{a}}
\epsilon
&
\begin{tikzpicture}[baseline=0]
\node at (0,0) (n1){$\NT{B}$};
\end{tikzpicture}
&\xrightarrow{\sym{b}}
\epsilon
\\
\begin{tikzpicture}[baseline=0]
\node at (0,0) (n1){$\NT{C}$};
\end{tikzpicture}
&\xrightarrow{\sym{c}}
\epsilon
&
\begin{tikzpicture}[baseline=0]
\node at (0,0) (n1){$\NT{D}$};
\end{tikzpicture}
&\xrightarrow{\sym{d}}
\epsilon
\end{aligned}$
&
\smash{$\begin{aligned}[t]
\begin{tikzpicture}[baseline=0]
\node(n1){$\NT{S}$};
\end{tikzpicture}
&\xRightarrow{\epsilon}
\begin{tikzpicture}[baseline=0]
\node(n1){$\NT{T}$};
\dnode{n1}{n2}{$\NT{L}_3$}
\end{tikzpicture}
\xRightarrow{\epsilon}
\begin{tikzpicture}[baseline=0]
\node(n1){$\NT{L}_1$};
\dnode{n1}{n2}{$\NT{T}$}
\dnode{n2}{n3}{$\NT{L}_2$}
\dnode{n3}{n4}{$\NT{L}_3$}
\end{tikzpicture}
\hspace{-6pt}\xRightarrow{\epsilon}
\begin{tikzpicture}[baseline=0]
\node(n1){$\NT{A}$};
\rnode{n1}{n2}{$\nt{X}$}
\rnode{n2}{n3}{$\NT{D}$}
\dnode{n2}{n4}{$\NT{T}$}
\dnode{n4}{n5}{$\NT{L}_2$}
\dnode{n5}{n6}{$\NT{L}_3$}
\end{tikzpicture}
\hspace{-6pt}\xRightarrow{\sym{a}}
\begin{tikzpicture}[baseline=0]
\node(n2){$\nt{X}$};
\rnode{n2}{n3}{$\NT{D}$}
\dnode{n2}{n4}{$\NT{T}$}
\dnode{n4}{n5}{$\NT{L}_2$}
\dnode{n5}{n6}{$\NT{L}_3$}
\end{tikzpicture}
\hspace{-6pt}\xRightarrow{\epsilon}
\begin{tikzpicture}[baseline=0]
\node(n2){$\nt{X}$};
\rnode{n2}{n3}{$\NT{D}$}
\dnode{n2}{n4}{$\NT{L}_1$}
\dnode{n4}{n5}{$\NT{T}$}
\dnode{n5}{n6}{$\NT{L}_2$}
\dnode{n6}{n7}{$\NT{L}_2$}
\dnode{n7}{n8}{$\NT{L}_3$}
\end{tikzpicture}
\\[3mm]
&\xRightarrow{\epsilon}
\begin{tikzpicture}[baseline=0]
\node(n2){$\nt{X}$};
\rnode{n2}{n3}{$\NT{D}$}
\dnode{n2}{n4}{$\NT{A}$}
\rnode{n4}{n5}{$\nt{X}$}
\rnode{n5}{n6}{$\NT{D}$}
\dnode{n5}{n7}{$\NT{T}$}
\dnode{n7}{n8}{$\NT{L}_2$}
\dnode{n8}{n9}{$\NT{L}_2$}
\dnode{n9}{n10}{$\NT{L}_3$}
\end{tikzpicture}
\hspace{-18pt}\xRightarrow{\sym{a}}
\begin{tikzpicture}[baseline=0]
\node(n2){$\nt{X}$};
\rnode{n2}{n3}{$\NT{D}$}
\dnode{n2}{n5}{$\nt{X}$}
\rnode{n5}{n6}{$\NT{D}$}
\dnode{n5}{n7}{$\NT{T}$}
\dnode{n7}{n8}{$\NT{L}_2$}
\dnode{n8}{n9}{$\NT{L}_2$}
\dnode{n9}{n10}{$\NT{L}_3$}
\end{tikzpicture}
\hspace{-18pt}\xRightarrow{\epsilon}
\begin{tikzpicture}[baseline=0]
\node(n2){$\nt{X}$};
\rnode{n2}{n3}{$\NT{D}$}
\dnode{n2}{n5}{$\nt{X}$}
\rnode{n5}{n6}{$\NT{D}$}
\dnode{n5}{n7}{$\nt{X}$}
\dnode{n7}{n8}{$\NT{L}_2$}
\dnode{n8}{n9}{$\NT{L}_2$}
\dnode{n9}{n10}{$\NT{L}_3$}
\end{tikzpicture}
\hspace{-18pt}\xRightarrow{\epsilon}
\begin{tikzpicture}[baseline=0]
\node(n2){$\nt{X}$};
\rnode{n2}{n3}{$\NT{D}$}
\dnode{n2}{n5}{$\nt{X}$}
\rnode{n5}{n6}{$\NT{D}$}
\dnode{n5}{n7}{$\nt{X}$}
\dnode{n7}{n7a}{$\NT{B}$}
\rnode{n7a}{n8}{$\nt{X}$}
\rnode{n8}{n8a}{$\NT{C}$}
\dnode{n8}{n9}{$\NT{L}_2$}
\dnode{n9}{n10}{$\NT{L}_3$}
\end{tikzpicture}
\hspace{-18pt}\xRightarrow{\sym{b}}
\begin{tikzpicture}[baseline=0]
\node(n2){$\nt{X}$};
\rnode{n2}{n3}{$\NT{D}$}
\dnode{n2}{n5}{$\nt{X}$}
\rnode{n5}{n6}{$\NT{D}$}
\dnode{n5}{n7}{$\nt{X}$}
\dnode{n7}{n8}{$\nt{X}$}
\rnode{n8}{n8a}{$\NT{C}$}
\dnode{n8}{n9}{$\NT{L}_2$}
\dnode{n9}{n10}{$\NT{L}_3$}
\end{tikzpicture}
\\[6mm]
&\xRightarrow{\epsilon}
\begin{tikzpicture}[baseline=0]
\node(n2){$\nt{X}$};
\rnode{n2}{n3}{$\NT{D}$}
\dnode{n2}{n5}{$\nt{X}$}
\rnode{n5}{n6}{$\NT{D}$}
\dnode{n5}{n7}{$\nt{X}$}
\dnode{n7}{n8}{$\nt{X}$}
\rnode{n8}{n8a}{$\NT{C}$}
\dnode{n8}{n9}{$\NT{B}$}
\rnode{n9}{n9a}{$\nt{X}$}
\rnode{n9a}{n9b}{$\NT{C}$}
\dnode{n9a}{n10}{$\NT{L}_3$}
\end{tikzpicture}
\hspace{-18pt}\xRightarrow{\sym{b}}
\begin{tikzpicture}[baseline=0]
\node(n2){$\nt{X}$};
\rnode{n2}{n3}{$\NT{D}$}
\dnode{n2}{n5}{$\nt{X}$}
\rnode{n5}{n6}{$\NT{D}$}
\dnode{n5}{n7}{$\nt{X}$}
\dnode{n7}{n8}{$\nt{X}$}
\rnode{n8}{n8a}{$\NT{C}$}
\dnode{n8}{n9a}{$\nt{X}$}
\rnode{n9a}{n9b}{$\NT{C}$}
\dnode{n9a}{n10}{$\NT{L}_3$}
\end{tikzpicture}
\hspace{-18pt}\xRightarrow{\epsilon}
\begin{tikzpicture}[baseline=0]
\node(n2){$\nt{X}$};
\rnode{n2}{n3}{$\NT{D}$}
\dnode{n2}{n5}{$\nt{X}$}
\rnode{n5}{n6}{$\NT{D}$}
\dnode{n5}{n7}{$\nt{X}$}
\dnode{n7}{n8}{$\nt{X}$}
\rnode{n8}{n8a}{$\NT{C}$}
\dnode{n8}{n9b}{$\NT{C}$}
\end{tikzpicture}
\hspace{-12pt}\xRightarrow{\sym{c}}
\begin{tikzpicture}[baseline=0]
\node(n2){$\nt{X}$};
\rnode{n2}{n3}{$\NT{D}$}
\dnode{n2}{n5}{$\nt{X}$}
\rnode{n5}{n6}{$\NT{D}$}
\dnode{n5}{n7}{$\nt{X}$}
\dnode{n7}{n8a}{$\NT{C}$}
\end{tikzpicture}
\hspace{-6pt}\xRightarrow{\sym{c}}
\begin{tikzpicture}[baseline=0]
\node(n2){$\nt{X}$};
\rnode{n2}{n3}{$\NT{D}$}
\dnode{n2}{n6}{$\NT{D}$}
\end{tikzpicture}
\xRightarrow{\sym{d}}
\begin{tikzpicture}[baseline=0]
\node{$\NT{D}$};
\end{tikzpicture}
\xRightarrow{\sym{d}}
\epsilon
\end{aligned}$} \\
\\[1ex]
PAA derivation tree for $\sym{aabbccdd}$ \\
{\tikzset{level distance=22pt}%
\def\axd{\begin{tikzpicture}[paa picture]\node(n1){$\NT{A}$};\rnode{n1}{n2}{$\nt{X}$}\rnode{n2}{n3}{$\NT{D}$}\fnode{n2}\end{tikzpicture}}
\def\bxc{\begin{tikzpicture}[paa picture]\node(n1){$\NT{B}$};\rnode{n1}{n2}{$\nt{X}$}\rnode{n2}{n3}{$\NT{C}$}\fnode{n2}\end{tikzpicture}}
\Tree [.$\NT{S}$
  [.$\NT{T}$
    [.$\NT{L}_1$ [.\axd{} \edge node[edge label] {$\sym{a}$}; [.$\NT{D}$ \edge node[edge label] {$\sym{d}$}; $\varepsilon$ ] ] ]
  [.$\NT{T}$
    [.$\NT{L}_1$ [.\axd{} \edge node[edge label] {$\sym{a}$}; [.$\NT{D}$ \edge node[edge label] {$\sym{d}$}; $\varepsilon$ ] ] ]
    [.$\NT{T}$ $\nt{X}$ ]
    [.$\NT{L}_2$ [.\bxc{} \edge node[edge label] {$\sym{b}$}; [.$\NT{C}$ \edge node[edge label] {$\sym{c}$}; $\varepsilon$ ] ] ]
  ]
    [.$\NT{L}_2$ [.\bxc{} \edge node[edge label] {$\sym{b}$}; [.$\NT{C}$ \edge node[edge label] {$\sym{c}$}; $\varepsilon$ ] ] ]
  ]
  [.$\NT{L}_3$ $\varepsilon$ ]
]}
\end{tabular}

\caption{Example PAA with run and derivation tree for $\sym{aabbccdd}$. To reduce clutter, 
symbols have been renamed. Symbol $\nt{X}$ is constant; other uppercase symbols are variable.}
\label{fig:paa_example}
\end{figure*}

A pushdown adjoining automaton (PAA) maintains a stack of symbols. The stack symbols are either \emph{variable} symbols, which can be rewritten, or \emph{constant} symbols, which cannot. Each stack symbol is augmented with another stack, written below it:
\begin{center}
\begin{tikzpicture}[paa picture]
\node(n1){$\ntvar{A}$};
\rnode{n1}{n2}{$\ntvar{B}$}
\rnode{n2}{n3}{$\ntvar{C}$}
\dnode{n2}{n4}{$\ntvar{D}$}
\rnode{n4}{n5}{$\ntvar{E}$}
\end{tikzpicture}
\medskip
\end{center}
Here, the top-level stack is $\ntvar{A}\ntvar{B}\ntvar{C}$, in which $\ntvar{B}$ is augmented with stack $\ntvar{D}\ntvar{E}$. The other symbols have empty stacks, which are not shown.

A PAA transition has a lhs which is a variable symbol
and a rhs which is a stack (as described above).
The rhs may have one symbol occurrence called the \emph{foot}; if so, both the lhs and the foot are written as
\begin{tikzpicture}[paa picture]\node(n){$\NTvar{X}$};\fnode{n}\end{tikzpicture}.

At each step, the PAA operates on the topmost stack that has a variable $\NTvar{X}$ on top. If it has a transition $\NTvar{X} \xrightarrow{\symvar{a}} \tree$, then it scans $\symvar{a}$, pops $\NTvar{X}$ and pushes $\tree$. If $\NTvar{X}$ had a stack attached, then the stack becomes attached to the foot of $\tree$.
Finally, any constant symbol that has an empty stack is deleted.

\begin{example}
The PAA for $\{\sym{a}^n\sym{b}^n\sym{c}^n\sym{d}^n\}$ is shown in \cref{fig:paa_example}, along with the run of this PAA on $\sym{aabbccdd}$.
\end{example}

The nested stacks of a PAA, described informally above, are treated formally as trees with a special constant symbol $\top$ at the root. However, we continue to draw them as shown above, without~$\top$.

\begin{defin}
A \defn{pushdown adjoining automaton} (PAA) is a tuple $\pushdown = (
\alphabet, \mathcal{V}, \mathcal{C}, \trans, 
\NTvar{S}
)$ where 
$\alphabet$, $\mathcal{V}$, and $\mathcal{C}$ are finite sets of 
\emph{input symbols}, \emph{variable stack symbols}, and \emph{constant stack symbols}, $\trans$ is a set of \emph{transitions} and 
$\NTvar{S} \in \mathcal{V}$ is the \emph{initial stack}.
The transitions are of the form $
\NTvar{X} \xrightarrow{\symvar{a}} 
\tree$ where $\NTvar{X} \in \mathcal{V}$ and $\tree \in \mathcal{T}(\mathcal{V} \cup \mathcal{C})$. 
The root of $\tree$ is $\top$, and at most one leaf in $\tree$ is designated the \emph{foot} and written as~\begin{tikzpicture}[paa picture]\node(n){$\NTvar{X}$};\fnode{n}\end{tikzpicture}.
\end{defin}

\begin{defin}
The path from the root to the foot of a PAA production's rhs is called its \defn{spine}. 
We say that a PAA production is \defn{spinal} if every rhs node is either on the spine or a child of a node on the spine.
A PAA is \defn{spinal} if all of its productions are spinal.
\end{defin}

\begin{defin}
If $\tree \in \mathcal{T}(\mathcal{V} \cup \mathcal{C})$ without constant leaf nodes, the \defn{top variable} of $\tree$ is defined as follows: If $\tree = \NTvar{X}$, then the top variable is $\NTvar{X}$. If $\tree = \begin{tikzpicture}[paa picture]\node(n){$\NTvar{X}$};\dnode{n}{n1}{$\tree_1 \cdots$}\end{tikzpicture}$ where $\ntvar{X} \in \mathcal{V}$, then the top variable is $\ntvar{X}$. If $\tree = \begin{tikzpicture}[paa picture]\node(n){$\ntvar{X}$};\dnode{n}{n1}{$\tree_1 \cdots$}\end{tikzpicture}$ where $\ntvar{X} \in \mathcal{C}$, then the top variable of $\tree$ is the top variable of $\tree_1$.
\end{defin}

\begin{defin}
If $\tree$ has top variable $\NTvar{X}$, and there is a transition $\atrans = (\NTvar{X} \xrightarrow{\symvar{a}} \rho)$, 
then we write $\tree \xRightarrow{\atrans} \tree'$ 
if $\tree'$ can be obtained from $\tree$ by replacing its top variable $\NTvar{X}$ with $\rho$ (sans $\top$), making the children of $\NTvar{X}$ the children of the foot of $\rho$, and deleting any constant leaf nodes.
\end{defin}


In a PAA derivation tree (see \cref{fig:paa_example} for an example), every node is a stack, written without its embedded stacks. 
If $\NTvar{X} \stackseq$ is such a node and $(\NTvar{X} \xrightarrow{\symvar{a}} \tree)$ is a transition, then the stacks of $\tree\stackseq$ are its children.

\begin{proposition} \label{thm:strong-paa}
A spinal PAA is \topeqJJ{} to a CFG $\control$ PDA.
\end{proposition}

\begin{proof} 
See \cref{app:d-strong-eq}; we give a brief sketch here, using the same three cases as in \cref{sec:control-derivations}.

Case (a) corresponds to the rule
\[
\NTvariii{A}{X}{} \xrightarrow{a} 
\varepsilon \]

Case (b) corresponds to the rule
\begin{equation*}
\tikzset{paa picture} \small
\begin{tikzpicture} \node(n1){$\NTvariii{A}{X}{Z}$}; \fnode{n1} \end{tikzpicture}
\xrightarrow{\varepsilon} 
\begin{tikzpicture}
\node(n1){$\NTvariii{S_2}{Y_1}{}$};
\rnode{n1}{n2}{${}\cdots{}$}
\rnode{n2}{n3}{$\NTvariii{S_2}{Y_{d-1}}{}$}
\rnode{n3}{n4}{$\,Z\,$}
\fnode{n4}
\rnode{n4}{n5}{$\NTvariii{S_2}{Y_{d+1}}{}$}
\rnode{n5}{n6}{${}\cdots{}$}
\rnode{n6}{n7}{$\NTvariii{S_2}{Y_k}{}$}
\end{tikzpicture}
\end{equation*}
where $\ntvar{Z}$ is constant.

Case (c) corresponds to the rules
{\tikzset{paa picture}
\begin{align*}
\begin{tikzpicture}
\node(n1){$\NTvariii{A}{X}{Z}$};
\fnode{n1}
\end{tikzpicture} 
&\xrightarrow{\varepsilon} 
\begin{tikzpicture}
\node(n1){$\NTvariii{B_1}{X}{Y_1}$};
\dnode{n1}{n2}{$\NTvariii{B_2}{Y_1}{Y_2}$}
\dotsnode{n2}{n3}{$\NTvariii{B_k}{Y_{k-1}}{Z}$}
\fnode{n3}
\end{tikzpicture} \\
\begin{tikzpicture}
\node(n1){$\NTvariii{A}{X}{}$};
\end{tikzpicture} 
&\xrightarrow{\varepsilon} 
\begin{tikzpicture}
\node(n1){$\NTvariii{B_1}{X}{Y_1}$};
\dnode{n1}{n2}{$\NTvariii{B_2}{Y_1}{Y_2}$}
\dotsnode{n2}{n3}{$\NTvariii{B_k}{Y_{k-1}}{}$}
\end{tikzpicture}
\\
\intertext{If $k=0$ then we must insert a constant symbol:}
\begin{tikzpicture}\node(n1){$\NTvariii{A}{X}{X}$}; \fnode{n1}\end{tikzpicture} &\xrightarrow{\varepsilon} 
\begin{tikzpicture}\node(n1){$\ntvar{X}$}; \fnode{n1}\end{tikzpicture} \qquad
\begin{tikzpicture}\node(n1){$\NTvariii{A}{X}{}$}; \end{tikzpicture} \xrightarrow{\varepsilon} 
\begin{tikzpicture}\node(n1){$\ntvar{X}$}; \end{tikzpicture} \tag*{\qedhere}
\end{align*}}
\end{proof}
\section{Conclusion}

We introduced new notions of equivalence between formalisms, \bijeqNN{} and \topeqNN{}, that allow for finer-grained comparisons than weak equivalence. By extending Weir's idea of control to PDAs, we obtained three new formalisms recognizing $\mathcal{L}_2$, all \bijeqJJ{} to Weir's original two-level grammar.
But by nuancing the idea of control to account for the difference between CFG derivations and derivation trees, we showed that they are not \topeqJJ. Instead, three of them are \topeqJJ{} to existing $\tal$ formalisms, namely TAG, LIG and EPDA, and the fourth is \topeqJJ{} to our new $\tal{}$ formalism, PAA.

\section*{Limitations}

We currently give separate definitions of the derivation trees of the four two-level formalisms. Ideally, one could give a general definition of derivation tree in a multi-level formalism, and then derive each particular case by plugging in the definition of derivation trees of (LD-)CFGs and (LD-)PDAs. We leave this generalization for future work.

The \topeqNN{} results only hold for spinal TAG and spinal PAA. However, full TAG and PAA are not \topeqJJ{} to spinal TAG and spinal PAA, and therefore they are not \topeqJJ{} to CFG $\control$ CFG and CFG  $\control$ PDA, either.

\section*{Ethics Statement}
The authors foresee no ethical concerns with the research presented in this paper.

\section*{Acknowledgements}
We would like to thank the anonymous reviewers for their comments and suggestions.
\david{to do}

\bibliography{custom}
\bibliographystyle{acl_natbib}
\appendix

\section{Context-Free Formal Systems} \label{app:context-free-fs}

\subsection{Context-Free Grammars}
\label{sec:cfg}

\begin{defin}
A \defn{context-free grammar} (CFG) is a tuple $\grammar = (\nonterm, \alphabet, \rules, \ntvar{S})$, where $\nonterm$ and $\alphabet$ are finite sets of \emph{nonterminal} and \emph{terminal symbols}, $\rules$ is a finite set of \emph{production rules}, and $\ntvar{S} \in \nonterm$ is a distinguished \emph{start symbol}. The production rules are of the form $\ntvar{A} \rightarrow \sent$, where $\ntvar{A} \in \nonterm$ and $\sent \in \kleene{(\nonterm \cup \alphabet)}$.
\end{defin}

\begin{defin}
A \defn{sentential form} of a CFG $\grammar$ is a sequence of nonterminal and terminal symbols $\sent \in \kleene{(\nonterm \cup \alphabet)}$.
If $p = (\ntvar{X}\rightarrow \sent) \in \rules$ and $\boldsymbol{\beta}_1, \boldsymbol{\beta}_2 \in \kleene{(\nonterm \cup \alphabet)}$, we write $\boldsymbol{\beta}_1 \ntvar{X} \boldsymbol{\beta}_2 \xRightarrow{p} \boldsymbol{\beta}_1 \sent \boldsymbol{\beta}_2$.
If
\[ \ntvar{S} \xRightarrow{p_1} \sent_1 \xRightarrow{p_2} \cdots \xRightarrow{p_n} \sent_n \]
where $\sent_n \in \kleene\alphabet$,
we say that $\grammar$ \defn{derives} $\sent_n$. The language generated by $\grammar$ is $\mathcal{L}(\grammar) = \{ \str \mid \text{$\grammar$ derives $\str$} \}$.
\end{defin}


\begin{defin} \label{def:cfg-derivation-tree}
The set of \defn{derivation trees} of a CFG is the smallest set defined as follows.
\begin{itemize}
\item If $\symvar{a} \in \alphabet$ then the tree consisting of a single node $\symvar{a}$ is an $\symvar{a}$-type derivation tree (whose yield is $\symvar{a}$).
\item If $(\ntvar{X} \rightarrow \ntvar{\beta_1} \cdots \ntvar{\beta_k}) \in \rules$ and, for $i=1, \ldots, k$, $\derivation_i$ is a $\beta_i$-type derivation tree, then the following is an $\ntvar{X}$-type derivation tree:
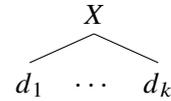
\[ \Tree[.$\ntvar{X}$ $\derivation_1$  \edge[draw=none]; $\cdots$ $\derivation_k$ ] \]
whose yield is $\yield(\derivation) = \yield(\derivation_1) \cdots \yield(\derivation_k)$.
\end{itemize}
A derivation tree of $\grammar$ is an $\ntvar{S}$-type derivation tree.
\end{defin}

\subsection{Pushdown Automata}
\label{sec:pda}

\begin{defin}
A \defn{configuration} of a PDA is a pair $(q,\stackseq)$, where $q\in\states$ is the current state and $\stackseq$ is the current contents of the stack.
\end{defin}

\begin{defin}
    If $(p, \ntvar{X} \boldsymbol\beta)$ and $(q, \stackseq \boldsymbol\beta)$ are configurations, and $\atrans = (p, \ntvar{X} \xrightarrow{\symvar{a}} q, \stackseq)$ is a transition, we write $(p,  \ntvar{X} \boldsymbol\beta) \xRightarrow{\atrans} (q, \stackseq \boldsymbol\beta)$. 
\end{defin}

\begin{defin}
A \defn{run} of a PDA is a sequence of transitions $\arun = (\atrans_1, \ldots, \atrans_n)$ such that there are configurations $(q_0, \stackseq_0), \ldots, (q_n, \stackseq_n)$ such that
\[(q_0, \stackseq_0) \xRightarrow{\atrans_1} \cdots \xRightarrow{\atrans_n} (q_n, \stackseq_n). \]

If each $\atrans_i$ scans $\symvar{a}_i$, then we say that the run scans $\symvar{a}_1 \cdots \symvar{a}_n$.

A run $\arun$ is called a \defn{pop computation} of $\ntvar{X}$ from $q_0$ to $q_n$ if $\stackseq_0 = \ntvar{X}\stackseq_n$, and for all $i<n$, $|\stackseq_i| \ge |\stackseq_0|$.

An \defn{accepting run} (or \defn{derivation}) is a pop computation of $\ntvar{S}$ from $q_\init$ to $q_\final$.
\end{defin}

\begin{defin}
Let $\pushdown$ be a PDA. If $\pushdown$ has an accepting run that scans $\str \in \kleene{\alphabet}$, we say that $\pushdown$ \defn{accepts} $\str$. The language \defn{recognized} by $\pushdown$ is $\mathcal{L}(\pushdown) = \{\str \in \kleene{\alphabet} \mid \text{$\pushdown$ accepts $\str$}\}$.
\end{defin}

\begin{defin} \label{def:pda-derivation-tree}
The set of \defn{derivation trees} of a PDA is the smallest set defined as follows.
\begin{itemize}
\item The tree consisting of a single node $(q_\final,\varepsilon)$ (which is the final configuration), is a $(q_\final,\varepsilon)$-type derivation tree (whose yield is~$\varepsilon$).
\item If $(q, \stackseq) \xRightarrow{\symvar{a}} (q', \stackseq')$ and $D'$ is a $(q', \stackseq')$-type derivation tree, then the following is a $(q, \stackseq)$-type derivation tree:
\[ D = \Tree [.{$(q,\stackseq)$} \edge node[edge label] {$\symvar{a}$}; {$D'$} ] \]
If the yield of $D'$ is $w'$, then the yield of $D$ is~$aw'$.
\end{itemize}
A derivation tree of $P$ is a $(q_\init, \ntvar{S})$-type derivation tree, where $(q_\init, \ntvar{S})$ is the initial configuration.
\end{defin}

The definitions of runs, accepting runs, and pop computation for LD-PDAs are analogous to those for PDAs.

\begin{proposition} \label{thm:onestatepda}
Every PDA is \topeqJJ{} to a PDA with one state.
\end{proposition}
\begin{proof}
Given a PDA 
\[\pushdown = (\states, \alphabet, \stackalphabet, \trans, (q_\init, \ntvar{S}), (q_\final, \varepsilon))\] 
construct the PDA
\begin{align*}
\pushdown' &= (\{q'\}, \alphabet, \stackalphabet', \trans', (q', \ntvar{S'}), (q', \varepsilon)) \\
\stackalphabet' &= \{X_{qr} \mid X \in \stackalphabet, q, r \in \states\} \\
\ntvar{S'} &= \ntvar{S}^{q_\init q_\final}
\end{align*}
and for every transition $(q, X \xrightarrow{a} r, Y_1 Y_2 \cdots Y_k) \in \trans$ with $k>0$, let $\trans'$ contain transitions $(q', X^{qs_k} \xrightarrow{a} q', Y_1^{rs_1} Y_2^{s_1s_2} \cdots Y_k^{s_{k-1}s_k})$ for all $s_1, \ldots, s_k \in \states$.
For every transition $(q, X \xrightarrow{a} r, \varepsilon) \in \trans$, let $\trans'$ contain transition $q', X^{qr} \xrightarrow{a} q', \varepsilon$.

We claim that $\pushdown$ and $\pushdown'$ are \topeqJJ. 
$(\Rightarrow)$ Given a derivation of $\pushdown$, apply the above construction to each of the transitions used. Although the construction generates many possible sequences of transitions, exactly one sequence is a well-formed derivation.
$(\Leftarrow)$ Given a derivation of $\pushdown'$, just apply the reverse construction to each of the transitions used.
\end{proof}
\section{$\tal{}$ Formalisms}\label{app:tal}

\subsection{Embedded Pushdown Automata}

To distinguish between the two types of stacks in an EPDA, we will call them the \emph{outer} stack and the \emph{inner} stacks.
A inner stack is written as $[\stackseq]$, $\stackseq \in \kleene\stackalphabet$ is written from top to bottom as in a PDA.
We write outer stacks as $\nestedstack=[\stackseq_1]\cdots [\stackseq_n]$, where $[\stackseq_1]$ is at the top, while $[\stackseq_n]$ is at the bottom.

\begin{defin}
A \defn{configuration} of an EPDA is a pair $(q,\nestedstack)$, where $q$ is the current state and $\nestedstack$ is the current contents of the outer stack.
\end{defin}

\begin{defin}
If an EPDA has a transition $\atrans = (q,[\NTvar{A}\rest]\xrightarrow{\symvar{a}}r,\epdastack{\nestedstack_1}{[\stackseq\rest]}{\nestedstack_2})$, then we write, for any $\boldsymbol{\Psi}\in \kleene{([\kleene{\stackalphabet}])}$ and  $\boldsymbol\beta \in \kleene{\stackalphabet}$,
\[(q, [\NTvar{A}\boldsymbol\beta]\,\boldsymbol\Psi)\xRightarrow{\atrans}(r,\nestedstack_1\,[\stackseq \boldsymbol\beta]\,\nestedstack_2\,\boldsymbol\Psi).\]
Similarly, if an EPDA has a transition $\atrans = (q,[\NTvar{A}]\xrightarrow{\symvar{a}}r,\varepsilon)$, then we write
\[(q, [\NTvar{A}]\,\boldsymbol\Psi) \xRightarrow{\atrans} (r, \boldsymbol\Psi). \]
\end{defin}

The definitions of scanning and non-scanning transitions, runs, and accepting runs of an EPDA, and EPDAs accepting strings and recognizing languages, are all analogous to those for PDAs.

\begin{defin}
    An EPDA is in \defn{\normalform{}} if its transitions have one of the following types
    \begin{align*}
        p,[\NTvar{A}\rest] &\xrightarrow{\varepsilon} q, [\NTvar{S}]\cdots [\rest] \cdots [\NTvar{S}], \\
        p,[\NTvar{A}\rest] &\xrightarrow{\varepsilon} p,  [\NTvar{B}_1 \cdots \NTvar{B}_k\rest], \\
        p, [\NTvar{A}] &\xrightarrow{\symvar{a}} q,\varepsilon.
    \end{align*}
\end{defin}


\subsection{Linear Indexed Grammars}

\begin{defin}
A \defn{sentential form} $\sent \in \kleene{(\nonterm[\kleene{\stackalphabet}] \cup \alphabet)}$ of a LIG is a sequence of nonterminal symbols, each augmented with a stack, and terminal symbols.
\end{defin}

\begin{defin}
If a LIG has a production $p = (\ligprod{X}{A}{\nestedstack_1}{Y}{\stackseq}{\nestedstack_2})$, then we write, for any $\boldsymbol{\Psi}_1, \boldsymbol{\Psi}_2 \in \kleene{(\nonterm[\kleene{\stackalphabet}] \cup \alphabet)}$ and  $\boldsymbol\beta \in \kleene{\stackalphabet}$,
\[\boldsymbol{\Psi}_1 \, \ntvar{X} [\NTvar{A} \boldsymbol\beta]\,\boldsymbol{\Psi}_2\xRightarrow{p}\boldsymbol{\Psi}_1\,\nestedstack_1\,\ntvar{Y} [\stackseq \boldsymbol\beta]\,\nestedstack_2\,\boldsymbol{\Psi}_2.\]

Similarly, if a LIG has a production $p = (\ntvar{X}[\NTvar{A}] \rightarrow \symvar{a})$, then we write
\[\boldsymbol{\Psi}_1\,\ntvar{X} [\NTvar{A}]\,\boldsymbol{\Psi}_2 \xRightarrow{p} \boldsymbol{\Psi}_1\,\symvar{a}\,\boldsymbol{\Psi}_2.\]
\end{defin}

A LIG derivation starts with the start symbol and initial stack $\ntvar{S}[\NTvar{S}]$. The definitions of LIG derivations, LIG derivation trees, LIGs accepting strings, and LIGs recognizing languages are all analogous to those for CFGs.

\begin{defin}
    A LIG is in \defn{\normalform{}} if its productions have one of the following types
    \begin{align*}
        \ntvar{X}[\NTvar{A}\rest] &\rightarrow \ntvar{Y}_1 [\NTvar{S}]\cdots \ntvar{Y}_{d} [\rest] \cdots \ntvar{Y}_k [\NTvar{S}], \\
        \ntvar{X}[\NTvar{A}\rest] &\rightarrow \ntvar{X} [\NTvar{B}_1 \cdots \NTvar{B}_k \rest], \\
        \ntvar{X}[\NTvar{A}] &\rightarrow \symvar{a}.
    \end{align*}
\end{defin}

\subsection{Tree-Adjoining Grammars}
\label{app:tag}

\begin{defin}
A tree $\beta$ with a foot node is \defn{adjoined} at an interior node $\eta$ by removing the children of $\eta$, replacing $\eta$ with $\beta$ and inserting the old children of $\eta$ as the children of the foot node of $\beta$. The old label of $\eta$ is thus lost. 

A tree $\alpha$ without a foot node is \defn{substituted} at a leaf node $\eta$ by replacing $\eta$ with $\alpha$. The old label of $\eta$ is thus lost.
\end{defin}

\begin{defin}
If $\atrans = (\NTvar{X} \rightarrow \beta)$ is a TAG production, $\alpha$ is a tree with a node $\eta$ labeled $\NTvar{X}$, and $\alpha'$ is the resulting of adjoining or substituting $\beta$ at $\eta$, we write $\alpha \xRightarrow{\atrans} \alpha'$.
\end{defin}

\begin{defin}
Let $\grammar$ be a TAG.
If
\begin{equation*}
    \NTvar{S} \xRightarrow{\atrans_1} \alpha_1 \Rightarrow \cdots \xRightarrow{\atrans_n} \alpha_n
\end{equation*}
where $\alpha_n$ does not have any variable nodes, then we say that $\grammar$ \defn{derives} $\alpha_n$.
\end{defin}

\begin{defin}
The \defn{tree language} of a TAG is the set of trees it derives.
Additionally, the \emph{yield} of a tree is the string obtained by concatenating the terminal symbols at the leaves of the tree, and
the \defn{string language} of a TAG is the set of strings yielded by its tree language.
\end{defin}

\begin{defin}
    A TAG is in \defn{\normalform} if its productions are of the types shown in \cref{fig:tag-rules-normal-form}.
    \begin{figure}
        \centering
        \footnotesize
        \begin{subfigure}{\linewidth}
        \[
            \NTvar{A} \rightarrow \hspace{-1.5em}
            \begin{tikzpicture}[baseline=0]
            \Tree [.$\ntvar{X}$ ${\NTvar{Y}_1}\subst$ \edge[draw=none]; $\mathclap{\cdots}$ ${\NTvar{Y}_{d-1}}\subst$ $\ntvar{Z}\foot$ ${\NTvar{Y}_{d+1}}\subst$ \edge[draw=none];  $\mathclap{\cdots}$ ${\NTvar{Y}_k}\subst$ ]
            \end{tikzpicture}
            \]
        \end{subfigure}

    \vspace*{3ex}
        
        \begin{subfigure}[t]{0.2\linewidth}
        \[
            \NTvar{A} \rightarrow
            \begin{tikzpicture}[baseline=0]
            \Tree [.${\NTvar{B}_1}$ [.${\NTvar{B}_2}$ \edge[dotted,very thick]; [.${\NTvar{B}_k}\foot$ ] ] ]
            \end{tikzpicture}
            \]
        \end{subfigure}
        \begin{subfigure}[t]{0.2\linewidth}
        \[
            \NTvar{A} \rightarrow
            \begin{tikzpicture}[baseline=0]
            \Tree [.${\NTvar{B}_1}$ [.${\NTvar{B}_2}$ \edge[dotted,very thick]; [.${\NTvar{B}_k}\subst$ ] ] ]
            \end{tikzpicture}
            \]
        \end{subfigure}
        \begin{subfigure}[t]{0.2\linewidth}
        \[
          \NTvar{A} \rightarrow \ntvar{X}\foot
        \]
        \end{subfigure}
        \begin{subfigure}[t]{0.2\linewidth}
        \[
            \NTvar{A} \rightarrow
            \begin{tikzpicture}[baseline=0]
            \Tree [.$\ntvar{X}$ $\symvar{a}$ ] 
            \end{tikzpicture} 
            \]
        \end{subfigure}
        \caption{Production rules of TAG in \normalform{}. Symbols $\ntvar{X}$ and $\ntvar{Z}$ are constant; other uppercase letters are variables.}
        \label{fig:tag-rules-normal-form}
    \end{figure}
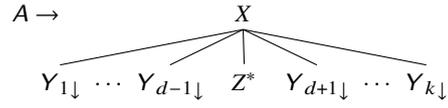
\end{defin}

\subsection{Pushdown Adjoining Automata}

\begin{defin}
    A PAA is in \defn{\normalform{}} if its transitions are of the types shown in \cref{fig:paa-rules-normal-form}. 

    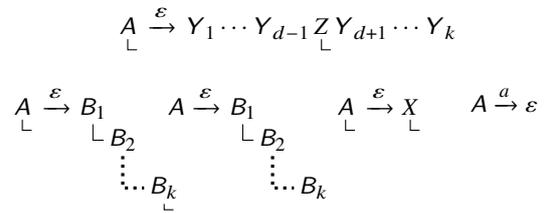
\begin{figure}
        \centering
        \begin{subfigure}[t]{\linewidth}
        \[\tikzset{paa picture} \footnotesize
        \begin{tikzpicture} \node(n1){$\NTvar{A}$}; \fnode{n1} \end{tikzpicture}
        \xrightarrow{\varepsilon} 
        \begin{tikzpicture}
        \node(n1){$\NTvar{Y}_1$};
        \rnode{n1}{n2}{${}\cdots{}$}
        \rnode{n2}{n3}{$\NTvar{Y}_{d-1}$}
        \rnode{n3}{n4}{$\,\ntvar{Z}\,$}
        \fnode{n4}
        \rnode{n4}{n5}{$\NTvar{Y}_{d+1}$}
        \rnode{n5}{n6}{${}\cdots{}$}
        \rnode{n6}{n7}{$\NTvar{Y}_k$}
        \end{tikzpicture}
        \]
        \end{subfigure}
        \begin{subfigure}[t]{0.25\linewidth}
        \[\tikzset{paa picture}\footnotesize
        \begin{tikzpicture}
        \node(n1){$\NTvar{A}$};
        \fnode{n1}
        \end{tikzpicture} 
        \xrightarrow{\varepsilon} 
        \begin{tikzpicture}
        \node(n1){$\NTvar{B_1}$};
        \dnode{n1}{n2}{$\NTvar{B_2}$}
        \dotsnode{n2}{n3}{$\NTvar{B_k}$}
        \fnode{n3}
        \end{tikzpicture}
        \]
        \end{subfigure}
        \begin{subfigure}[t]{0.25\linewidth}
        \[\tikzset{paa picture}\footnotesize
        \begin{tikzpicture}
        \node(n1){$\NTvar{A}$};
        \end{tikzpicture} 
        \xrightarrow{\varepsilon} 
        \begin{tikzpicture}
        \node(n1){$\NTvar{B_1}$};
        \dnode{n1}{n2}{$\NTvar{B_2}$}
        \dotsnode{n2}{n3}{$\NTvar{B_k}$}
        \end{tikzpicture}
        \]
        \end{subfigure}
        \begin{subfigure}[t]{0.2\linewidth}
        \[\tikzset{paa picture}\footnotesize
        \begin{tikzpicture}\node(n1){$\NTvar{A}$}; \fnode{n1}\end{tikzpicture} \xrightarrow{\varepsilon} 
        \begin{tikzpicture}\node(n1){$\ntvar{X}$}; \fnode{n1}\end{tikzpicture}
        \] 
        \end{subfigure}
        \begin{subfigure}[t]{0.2\linewidth}
        \footnotesize
        \[
        \NTvar{A} \xrightarrow{\symvar{a}} 
        \varepsilon \]
        \end{subfigure}
        \caption{Transitions of PAA in \normalform{}. Symbols $\ntvar{X}$ and $\ntvar{Z}$ are constant; other uppercase letters are variables.}
        \label{fig:paa-rules-normal-form}
    \end{figure}
\end{defin}
\section{Derivation Trees of Two-Level Formalisms}\label{app:control-formalisms}

In \cref{sec:control-derivations} we showed what the derivation trees of four two-level formalisms look like when both the controller and the controllee are in \normalform{}. In this section, we give details on what the derivation trees look like for general controller CFGs/PDAs and controllee LD-CFGs/LD-PDAs.

\paragraph{PDA $\control$ PDA}

The \normalform{} and general form of PDAs are not so different, so moving to general PDAs is straightforward. We still distinguish between the cases where the controller transition is (a,b) scanning or (c) non-scanning, and if the former, whether the controllee transition has (a) no distinguished symbol or (b) a distinguished symbol.

(a) If a derivation node is labeled $q,\ntvar{X}[s, \NTvar{A}\boldsymbol\beta]\,\nestedstack$, 
and $s, \NTvar{A} \xrightarrow{\Symvar{\ell}} q_\final, \varepsilon$ is a controller transition,
and $\Symvar{\ell}$ is a controllee transition $q, \ntvar{X} \xrightarrow{\symvar{a}} r, \ntvar{Y_1} \cdots \ntvar{Y}_k$ (with no distinguished symbol),
\begin{center}
\begin{tikzpicture} \scriptsize
\Tree [.$q,{\ntvar{X}[s,\NTvar{A}]\,\nestedstack}$ 
\edge node[edge label] {$\symvar{a}$};
${r,\ntvar{Y}_1[q_\init,\NTvar{S_2}] \cdots \ntvar{Y}_k[q_\init,\NTvar{S_2}]\,\nestedstack}$ ]
\end{tikzpicture}
\end{center}

(b) If $s, \NTvar{A} \xrightarrow{\Symvar{\ell}} t, \stackseq$ is a controller transition,
and $\Symvar{\ell}$ is a controllee transition $q, \ntvar{X} \xrightarrow{\symvar{a}} r, \ntvar{Y_1} \cdots \ntvar{Y_{d-1}}\,\dist{\ntvar{Z}}\,\ntvar{Y_{d+1}} \cdots \ntvar{Y}_k$, 
then the node can have child
\begin{center}
\begin{tikzpicture} \scriptsize
\Tree [.$q,{\ntvar{X}[s,\NTvar{A}\boldsymbol\beta]\,\nestedstack}$ 
\edge node[edge label] {$\symvar{a}$};
${r,\ntvar{Y}_1[q_\init,\NTvar{S_2}] \cdots \ntvar{Y}_{d-1}[q_\init,\NTvar{S_2}]\,\ntvar{Z}[t,\stackseq\boldsymbol\beta]\,\ntvar{Y}_{d+1}[q_\init,\NTvar{S_2}] \cdots \ntvar{Y}_k[q_\init,\NTvar{S_2}]\,\nestedstack}$ ]
\end{tikzpicture}
\end{center}

Case (c) stays the same.

\paragraph{PDA $\control$ CFG} Generalizing to a PDA controller is also not very difficult here.

(a) If a derivation node is labeled $\ntvar{X}[q,\NTvar{A}]$, 
and $q,\NTvar{A} \xrightarrow{\Symvar{\ell}} q_\final,\varepsilon$ is a controller transition,
and $\Symvar{\ell}$ is a controllee production $\ntvar{X} \rightarrow \upsilon_1 \cdots \upsilon_k$, where each $\upsilon_i$ is either a terminal or non-distinguished nonterminal symbol,
then the node can have children as follows. For each $i$,
\begin{itemize}
\item If $\upsilon_i$ is a terminal $\symvar{a}$, then there is a child $\symvar{a}$.
\item If $\upsilon_i$ is a non-distinguished nonterminal $\ntvar{Y}$, then there is a child $\ntvar{Y}[q_\init, \NTvar{S}_2]$.
\end{itemize}

(b) If a derivation node is labeled $\ntvar{X}[q,\NTvar{A} \boldsymbol\beta]$, 
and $q,\NTvar{A} \xrightarrow{\Symvar{\ell}} r,\stackseq$ is a controller transition,
and $\Symvar{\ell}$ is a controllee production $\ntvar{X} \rightarrow \upsilon_1 \cdots \upsilon_k$, where each $\upsilon_i$ is either a terminal, nonterminal, or distinguished nonterminal symbol,
then the node can have children as follows. For each $i$,
\begin{itemize}
\item If $\upsilon_i$ is a terminal or non-distinguished nonterminal, then there is a child as in case (a).
\item If $\upsilon_i$ is a distinguished nonterminal $\ntvar{Z}$, then there is a child $\symvar{Z}[r,\stackseq\boldsymbol\beta]$.
\end{itemize}

Case (c) stays the same.

\paragraph{CFG $\control$ CFG}
Generalizing to a CFG controller is more complicated because a production can now yield multiple terminal symbols, which causes the controllee to use multiple productions in a single derivation step.

Derivation nodes are labeled $\NTvariii{A}{X}{Z}$ where $\NTvar{A}$ is a controller nonterminal, and $X$ and $Z$ are controllee nonterminals or $\bot$. We write $\NTvariii{A}{X}{\bot}$ as $\NTvariii{A}{X}{}$.

If a derivation node is labeled $\NTvariii{A}{X}{Z}$, and $\NTvar{A} \rightarrow \beta_1 \cdots \beta_k$ is a controller production, where each $\beta_i$ is either a nonterminal or a terminal symbol, then the derivation node can have children as follows. 
Let $\ntvar{Y_0} = \ntvar{X}$, $\ntvar{Y_k} = \ntvar{Z}$, and $\ntvar{Y_1}, \ldots, \ntvar{Y_{k-1}}$ be any controllee nonterminals.
For each $i$:
\begin{enumerate}
\item[(a)] If $\beta_i$ is a terminal $\Symvar{\ell}$ that names a controllee production $\ntvar{Y_{i-1}} \rightarrow \upsilon_1 \cdots \upsilon_m$, where each $\upsilon_j$ is a terminal or non-distinguished nonterminal (and $\ntvar{Y_i} = \bot$), then for each $j$,
\begin{itemize}
\item If $\upsilon_j$ is a terminal $\symvar{a}$, then there is a child~$\symvar{a}$.
\item If $\upsilon_j$ is a non-distinguished nonterminal $\ntvar{Y}$, then there is a child $\ntvar{Y}[\NTvar{S_2}]$.
\end{itemize}
\item[(b)] If $\beta_i$ is a terminal $\Symvar{\ell}$ that names a controllee production $\ntvar{Y_{i-1}} \rightarrow \upsilon_1 \cdots \upsilon_m$, where each $\upsilon_j$ is a terminal, nonterminal or distinguished nonterminal (and $\ntvar{Y_i} \neq \bot$), then for each $j$,
\begin{itemize}
\item If $\upsilon_j$ is a terminal or non-distinguished nonterminal, then there is a child as in case (a).
\item If $\upsilon_j$ is a distinguished nonterminal, it must be $\dist{\ntvar{Y}_i}$ (and $\ntvar{Y}_i \neq \bot$).
\end{itemize}
\item[(c)] If $\beta_i$ is a nonterminal $\NTvar{B}$, 
then there is a child $\NTvariii{B}{Y_{i-1}}{Y_i}$.
\end{enumerate}

\paragraph{CFG $\control$ PDA}

Derivation nodes are labeled $(q, \nestedstack)$, where $q$ is a state and $\nestedstack$ is a string of symbols of the form $\NTvariii{A}{X}{Z}$ (as above, for CFG $\control$ CFG) or $*$.

If a derivation node is labeled $q,\maybestar\NTvariii{A}{X}{Z}\,\nestedstack$
(where $\maybestar$ denotes either the presence or absence of a $*$),
and $\NTvar{A} \rightarrow \beta_1 \cdots \beta_k$ is a controller production,
where each $\beta_i$ is either a nonterminal or terminal symbol,
then the derivation node can have children as follows.
Let $\ntvar{Y_0} = \ntvar{X}$, $\ntvar{Y_k} = \ntvar{Z}$, and $\ntvar{Y_1}, \ldots, \ntvar{Y_{k-1}}$ be any controllee nonterminals. For each $i$:
\begin{itemize}
\item[(a)] If $\beta_i$ is a terminal $\Symvar{\ell}$ that names a controllee transition $q, \ntvar{Y_{i-1}} \xrightarrow{\symvar{a}} q, \ntvar{W_1} \cdots \ntvar{W_k}$ (with no distinguished symbol and $\ntvar{Y_i} = \bot$), then there is a child
\begin{center} \small \vspace*{-1ex}
\begin{tikzpicture}
\Tree [.{} \edge node[edge label] {$\symvar{a}$};
${q,\NTvariii{S_2}{W_1}{} \cdots \NTvariii{S_2}{W_k}{}}$ ]
\end{tikzpicture}
\end{center}
\item[(b)] If $\beta_i$ is a terminal $\Symvar{\ell}$ that names a controllee transition $q, \ntvar{Y_{i-1}} \xrightarrow{\symvar{a}} q, \ntvar{W_1} \cdots \ntvar{W_{d-1}}\,\dist{\ntvar{Y_i}}\,\ntvar{W_{d+1}} \cdots \ntvar{W_k}$ (and $\ntvar{Y_i} \neq \bot$),  then there is a child
\begin{center} \small \vspace*{-1ex}
\begin{tikzpicture}
\Tree [.{} \edge node[edge label] {$\symvar{a}$};
${q,\NTvariii{S_2}{W_1}{} \cdots \NTvariii{S_2}{W_{d-1}}{} * \NTvariii{S_2}{W_{d+1}}{} \cdots \NTvariii{S_2}{W_k}{}}$ ]
\end{tikzpicture}
\end{center}
\item[(c)] If $\beta_i$ is a nonterminal $\NTvar{B}$, then there is a child 
\begin{center} \small \vspace*{-1ex}
\begin{tikzpicture}
\Tree [.{} \edge node[edge label] {$\varepsilon$};
$q,\NTvariii{B}{Y_{i-1}}{Y_i}$ ]
\end{tikzpicture}
\end{center}
\end{itemize}
Finally, append $\nestedstack$ to the leftmost child.

\section{\BijEqNN{} Proofs}\label{app:d-weak-eq}

\begin{proposition} \label{proposition:cfg-pda-dweak-eq}
CFG and PDA are \bijeqJJ. 
\end{proposition}

\begin{proof}
    $(\Rightarrow)$ Let $\grammar = \left(\nonterm, \alphabet, \rules, \NTvar{S} \right)$ be a CFG in \normalform{}. We define the PDA in \normalform{} ${\pushdown}_{\grammar} = \left(\{q\}, \alphabet, \nonterm, \trans, \left(q, \NTvar{S} \right), \left(q, \varepsilon \right) \right)$, where 
    \begin{align*}
        \trans &= \{ q, \NTvar{A}\xrightarrow{\varepsilon} q, \NTvar{B}_1 \cdots \NTvar{B}_k \mid \NTvar{A} \rightarrow \NTvar{B}_1 \cdots \NTvar{B}_k \in \rules \} \\
        &\quad \cup \{q, \NTvar{A}\xrightarrow{\Symvar{a}} q, \varepsilon \mid \NTvar{A} \rightarrow \Symvar{a} \in \rules \}.
    \end{align*}
    We prove by induction on the height $h$ of the derivation tree that for each $\NTvar{A}$-type derivation tree $\derivation$ of $\grammar$ that yields $w$, $\pushdown_\grammar$ has a pop computation of $\NTvar{A}$ that scans $w$.

    \textbf{Base case:} When the height is $h=2$, it must hold that $\grammar$ has a production $\NTvar{A}\rightarrow \Symvar{a}$. The yield of the derivation is $\Symvar{a}$. By construction, $\pushdown_\grammar$ has the transition $q, \NTvar{A}\xrightarrow{\Symvar{a}} q, \varepsilon$, which is a pop computation of $\NTvar{A}$ that scans $\Symvar{a}$. 
    
    \textbf{Inductive Step:} We assume that the statement holds for each $\NTvar{A}$-type derivation of height up to $h$ and we assume that $\derivation$ is a derivation tree of height $h+1$ yielding $w = w_1 w_2 \cdots w_k$, which has root $\NTvar{A}$ and children $\NTvar{B}_1, \NTvar{B}_2, \ldots, \NTvar{B}_k$. We assume that the $\NTvar{B}_1, \NTvar{B}_2, \ldots, \NTvar{B}_k$-type derivations have yield $w_1, w_2, \ldots, w_k$. This implies that $\grammar$ has the production $\NTvar{A}\rightarrow \NTvar{B_1} \NTvar{B_2} \cdots \NTvar{B_k}$ and $\pushdown_\grammar$ has the transition $q, \NTvar{A} \xrightarrow{\varepsilon} q, \NTvar{B_1} \NTvar{B_2} \cdots \NTvar{B_k}$. By the inductive hypothesis, $\pushdown_\grammar$ has pop computations (from state $q$ to state $q$) of $\NTvar{B_1}, \NTvar{B_2}, \ldots, \NTvar{B_k}$ scanning $w_1, w_2, \ldots, w_k$, thus it has a pop computation of $\NTvar{A}$ that yields $w$.
    
    Therefore, there is a yield-preserving bijection between the $\NTvar{S}$-type derivations of $\grammar$ and $\pushdown_\grammar$'s pop computations of $\NTvar{S}$ (accepting runs).

    $(\Leftarrow)$ Let $\pushdown = (\states, \alphabet, \stackalphabet, (q_\init, \NTvar{S}), (q_\final, \varepsilon) )$ be a PDA in \normalform{}. We define the CFG in \normalform{} 
    \begin{align*}
    \grammar_{\pushdown} &= (\nonterm, \alphabet, \rules, \NTvar{S}') \\
    \nonterm &= \{ \NTvar{A}^{pq} \mid \NTvar{A} \in \stackalphabet, p,q \in \states \} \\
    \NTvar{S}' &= \NTvar{S}^{q_\init q_\final}
    \end{align*}
    and $\rules$ is constructed as follows:
    \begin{itemize}
    \item For each transition $(p, \NTvar{A} \xrightarrow{\varepsilon} q, \NTvar{B}_1 \cdots \NTvar{B}_k) \in \trans$, and for each $r_1, r_2, \ldots, r_k \in \states$, include production $\NTvar{A}_{p r_k} \rightarrow \NTvar{B}_1^{q r_1} \NTvar{B}_2^{r_1 r_2} \cdots \NTvar{B}_k^{r_{k-1} r_k}$.
    \item For each transition $(p, \NTvar{A} \xrightarrow{\Symvar{a}} q, \varepsilon) \in \trans$, include production $\NTvar{A}^{pq} \rightarrow \Symvar{a}$.
    \end{itemize}

    We prove that for each pop computation of $\NTvar{A}$ of $\pushdown$ from $p$ to $q$ that scans $w$, there is a $\NTvar{A}^{pq}$-type derivation of $\grammar_\pushdown$ with yield $w$. We prove the statement by induction on the length $l$ of the pop computation. 

    \textbf{Base case:} When the pop computation has length $l=1$, it must consist of a single transition $p, \NTvar{S} \xrightarrow{\Symvar{a}}q, \varepsilon$. It scans the string $\Symvar{a}$. By construction, $\grammar_{\pushdown}$ has a production $\NTvar{S}^{pq} \rightarrow \Symvar{a}$, thus it has a derivation yielding $\Symvar{a}$.
    
    \textbf{Inductive Step:} We assume that the statement holds for all pop computations of length up to $l$ and $\arun$ is a pop computation of $\NTvar{A}$ such that its first transition is $p, \NTvar{A}\xrightarrow{\varepsilon} q, \NTvar{B_1} \NTvar{B_2} \cdots \NTvar{B_k}$, followed by pop computations of $\NTvar{B_1}, \NTvar{B_2}, \ldots, \NTvar{B_k}$ from $p$ to $r_1$, $r_1$ to $r_2$, $\ldots$, and $r_{k-1}$ to $r_k$, respectively. We assume that the pop computations of $\NTvar{B_1}, \NTvar{B_2}, \ldots, \NTvar{B_k}$ scan $w_1, w_2, \ldots, w_k$, therefore $\arun$ scans $w = w_1 w_2 \cdots w_k$. By construction, $\grammar_\pushdown$ has a production $\NTvar{A}^{pq}\rightarrow \NTvar{B}_1^{p r_1} \NTvar{B}_2^{r_1 r_2} \cdots \NTvar{B}_k^{r_{k-1} r_k}$. By the inductive hypothesis $\grammar_\pushdown$ has $\NTvar{B}_1^{p r_1} \NTvar{B}_2^{r_1 r_2} \cdots \NTvar{B}_k^{r_{k-1} r_k}$-type derivations yielding $w_1 w_2 \cdots w_k$, thus it also has a $\NTvar{A}^{p r_k}$-type derivation yielding $w$. 
    
    Therefore, there is a yield-preserving bijection between the accepting runs of $\pushdown$ (pop computations of $\NTvar{S}$ from $q_\init$ to $q_\final$) and derivation trees of $\grammar_\pushdown$.
\end{proof}

\begin{proposition}\label{prop:ldcfg-ldpda-weak}
    LD-CFG and LD-PDA are \bijeqJJ, in the sense that
    for any LD-CFG $\grammar$, there is an LD-PDA $\pushdown_\grammar$ such that for any controller $\system$, $\system \control \grammar$ and $\system \control 
    \pushdown_\grammar$ are \bijeqJJ; 
    conversely, for any any LD-PDA $\pushdown$, there is an LD-PDA $\grammar_\pushdown$ such that for any controller $\system$, $\system \control \pushdown$ and $\system \control 
    \grammar_\pushdown$ are \bijeqJJ.
\end{proposition}

\begin{proof}
    $(\Rightarrow)$ Let $\grammar = \left( \nonterm, \alphabet, \labelset, \rules, \ntvar{S} \right)$ be an LD-CFG in \normalform{}. We define the LD-PDA \[\pushdown_{\grammar} = \left(\{q\}, \alphabet, \nonterm, \labelset, \trans, \left(q, \ntvar{S}\right), \left(q, \varepsilon \right) \right)\] where $\trans$ is constructed as follows:
    \begin{itemize}
    \item For each $(\Symvar{\ell}\colon \NTvar{A} \rightarrow \NTvar{B}_1 \cdots \dist{\NTvar{B}}_d \cdots \NTvar{B}_k)$ in $\rules$, include transition $\Symvar{\ell}\colon q, \NTvar{A}\xrightarrow{\varepsilon} q, \NTvar{B}_1 \cdots \dist{\NTvar{B}}_d \cdots \NTvar{B}_k$.
    \item For each $(\Symvar{\ell}\colon \NTvar{A} \rightarrow \Symvar{a}) \in \rules$, include transition $\Symvar{\ell}\colon q, \NTvar{A}\xrightarrow{\Symvar{a}} q, \varepsilon$. 
    \end{itemize}
    The rest of the proof is similar to the proof of \cref{proposition:cfg-pda-dweak-eq}. The only addition is to observe that $\pushdown_\grammar$ uses each control word the same number of times that $\grammar$ does, ensuring that the derivations of $\system$, $\system \control \grammar$ and $\system \control 
    \pushdown_\grammar$ are in one-to-one correspondence.\david{If I defined d-weak equivalence correctly above, I think the proof technically needs to be a little more involved than this, but maybe no one is willing to work it out! \response{david} I added one sentence as a halfhearted attempt}

    $(\Leftarrow)$ Let $\pushdown = \left(\states, \alphabet, \stackalphabet, \labelset, \left(q_\init, \ntvar{S} \right), \left(q_\final, \varepsilon \right) \right)$ be an LD-PDA in \normalform{}. We define the LD-CFG in \normalform{} 
    \begin{align*}
    \grammar_{\pushdown} &= (\nonterm, \alphabet, \rules, \ntvar{S}') \\
    \nonterm &= \{ \NTvar{A}^{pq} \mid \NTvar{A} \in \stackalphabet, p,q \in \states \} \\
    \ntvar{S}' &= \ntvar{S}^{q_\init q_\final}
    \end{align*}
    and $\rules$ is constructed as follows:
    \begin{itemize}
    \item For each $\Symvar{\ell}\colon p, \NTvar{A} \xrightarrow{\varepsilon} q, \NTvar{B}_1 \cdots \dist{\NTvar{B}}_d \cdots \NTvar{B}_k$ in $\trans$, and for each $r_1, r_2, \ldots, r_k \in \states$, include production $\Symvar{\ell}\colon \NTvar{A}^{p r_k} \rightarrow \NTvar{B}_1^{q r_1} \NTvar{B}_2^{r_1 r_2} \cdots \dist{\NTvar{B}}_d^{r_{d-1} r_d} \cdots \NTvar{B}_k^{r_{k-1} r_k}$.
    \item For each transition $(\Symvar{\ell}\colon p, \NTvar{A} \xrightarrow{\Symvar{a}} q, \varepsilon) \in \trans$, include production $\Symvar{\ell}\colon \NTvar{A}^{pq} \rightarrow \Symvar{a}$.
    \end{itemize}
    Again, the rest of the proof is similar to the proof of \cref{proposition:cfg-pda-dweak-eq}.
\end{proof}

\begin{repproposition}{thm:weak}
CFG $\control$ CFG, PDA $\control$ CFG, CFG $\control$ PDA and PDA $\control$ PDA are \bijeqJJ{}.
\end{repproposition}

\begin{proof}
    When two two-level formalisms have the same controllee but one has a controller CFG and the other has a controller PDA, it is sufficient to define controllers that are \bijeqJJ{}. The proof of \cref{proposition:cfg-pda-dweak-eq} shows that this is possible and how the controllers can be defined. If there is a bijection between the derivations of the CFG and PDA for each control word, the rules of the controllee will be applied in the same order in both two-level formalisms, thus generating the same set of strings with an equal number of derivations. 

    When two two-level formalisms have the same controller but different controllees, the controllees must consume the control words in the same order and the same number of times. Indeed, the proof of \cref{prop:ldcfg-ldpda-weak} shows that this is possible and how the controllees can be defined.
    
\end{proof}
\section{\TopEqNN{} Proofs}\label{app:d-strong-eq}

We assume for simplicity that the controller CFGs/PDAs and the LD-CFGs/LD-PDAs are in \normalform{} in this section. However, the proofs of equivalence also hold in the general case. 

\begin{repproposition}{thm:strong-epda}
EPDA and PDA $\control$ PDA are \topeqJJ{}.
\end{repproposition}

\begin{proof}

$(\Rightarrow)$ Let \[\pushdown = \left(\states, \alphabet, \stackalphabet, \trans, (q_\init, [\NTvar{S}]), (q_\final, \varepsilon) \right)\] be an EPDA in \normalform{}. Construct the controllee LD-PDA and controller PDA
\begin{align*}
\pushdown_1 &= (\states, \alphabet, \{ \ntvar{X} \}, \labelset, \trans_1, (q_\init, \ntvar{X}), (q_\final, \varepsilon)) \\
\pushdown_2 &= (\{q\}, \labelset, \stackalphabet, \trans_2, (q, \NTvar{S}), (q, \varepsilon))
\end{align*}
where the set of labels $\labelset$ and the sets of transitions $\trans_1$ and $\trans_2$ are constructed as follows:
\begin{itemize}
    \item For each transition $(p, [\NTvar{A}\rest]\xrightarrow{\varepsilon}p, [\stackseq \rest])\in \trans$, 
    let $\labelset$ contain a fresh label $\ell$,
    let $\trans_2$ contain the transition $q, \NTvar{A}\xrightarrow{\ell}q, \stackseq$,
    and let $\trans_1$ contain the transition $\ell \colon p, \ntvar{X} \xrightarrow{\varepsilon} p, \ntvar{X}$.
    \david{what about $p$? \response{alexandra} $p$ is the controllee state and is different from the single state of the controller $q$. we don't allow the controllee to change state when only taking a transition of the controller. although now that i think about it, I think we could with some state annotations. \response{david} right, otherwise the PDA-PDA will allow some moves that the EPDA didnt, right? \response{alexandra} yes \response{david} Does that fix it? This is similar to a trick that was used in the TAG proof.}
    \item For each transition $(p,[\NTvar{A}\rest]\xrightarrow{\varepsilon} r, [\NTvar{S}]\ldots [\rest] \ldots [\NTvar{S}])$ in $\trans$, add a new label $\Symvar{\ell}$ to $\labelset$, the transition $q, \NTvar{A} \xrightarrow{\Symvar{\ell}}q, \varepsilon$ to $\trans_2$ and the transition $\Symvar{\ell}\colon p,\ntvar{X} \xrightarrow{\varepsilon} r,\ntvar{X} \cdots \dist{\ntvar{X}} \cdots \ntvar{X}$ to $\trans_1$.
    \item For each transition $(p, [\NTvar{A}] \xrightarrow{\symvar{a}} r, \varepsilon) \in \trans$, add a new label $\Symvar{\ell}$ to $\labelset$, the transition $q, \NTvar{A} \xrightarrow{\Symvar{\ell}} q, \varepsilon$ to $\trans_2$ and the transition $\Symvar{\ell}\colon p,\ntvar{X} \xrightarrow{\symvar{a}} r, \varepsilon$ to $\trans_1$. 
\end{itemize}

$(\Leftarrow)$ 
We are given $\pushdown_2 \control \pushdown_1$, and by \cref{thm:onestatepda} we can assume without loss of generality that both $\pushdown_1$ and $\pushdown_2$ have only one state. Thus
\begin{align*}
\pushdown_1 &= (\{q\}, \alphabet, \stackalphabet_1, \labelset, \trans, (q, \ntvar{S}_1), (q, \varepsilon)) \\
\pushdown_2 &= (\{q\}, \labelset, \stackalphabet_2, \trans_2, (q, \NTvar{S}_2), (q, \varepsilon))
\end{align*}
and we construct the EPDA 
\begin{align*}
\pushdown' &= (\{q\}, \alphabet, \stackalphabet', \trans', (q, [\NTvariii{S_2}{S_1}{}]), (q, \varepsilon)) \\
\stackalphabet' &= \{ \NTvariii{A}{X}{Y} \mid \NTvar{A} \in \stackalphabet_2, \ntvar{X}, \ntvar{Y} \in \stackalphabet_1 \} \\
&\quad \cup \{ \NTvariii{A}{X}{} \mid \NTvar{A} \in \stackalphabet_2, \ntvar{X} \in \stackalphabet_1 \}.
\end{align*}
The set $\trans'$ is constructed as follows:
\begin{itemize}
    \item For each transition $(q, \NTvar{A} \xrightarrow{\varepsilon} q, \NTvar{B_1}\cdots \NTvar{B_k}) \in \trans_2$ where $k>0$, and for each $\ntvar{X_0}, \ldots, \ntvar{X_k} \in \stackalphabet_1$, let $\trans'$ contain the transition $q, [\NTvariii{A}{X_0}{X_k}\rest] \xrightarrow{\varepsilon} q, [\NTvariii{B_1}{X_0}{X_1} \cdots \NTvariii{B_k}{X_{k-1}}{X_k} \rest]$.
    \item For each transition $(q, \NTvar{A} \xrightarrow{\varepsilon} q, \varepsilon) \in \trans_2$, and for each $\ntvar{X} \in \stackalphabet_1$, let $\trans'$ contain the transition $q, [\NTvariii{A}{X}{X}\rest] \xrightarrow{\varepsilon} q, [\rest]$.

    \item For each pair of transitions $(q, \NTvar{A} \xrightarrow{\Symvar{\ell}} q, \varepsilon) \in \trans_2$ and $(\Symvar{\ell} \colon q, \ntvar{X} \xrightarrow{\varepsilon} q, \ntvar{Y_1}\cdots \dist{\ntvar{Y}}_d\cdots \ntvar{Y_k}) \in \trans_1$, let $\trans'$ contain the transition $q, [\NTvariii{A}{X}{Y_d} \rest]\xrightarrow{\varepsilon} q, [\NTvariii{S_2}{Y_1}{}]\cdots [\rest] \cdots [\NTvariii{S_2}{Y_k}{}]$ (where the $[\rest]$ is in the $d$-th position).
    \item For each pair of transitions $(q, \NTvar{A} \xrightarrow{\Symvar{\ell}} q, \varepsilon) \in \trans_2$ and $(\Symvar{\ell} \colon q, \ntvar{X} \xrightarrow{\symvar{a}} q, \varepsilon) \in \trans_1$, let $\trans'$ contain the transition $q, [\NTvariii{A}{X}{}] \xrightarrow{\symvar{a}} q, \varepsilon$. \qedhere
\end{itemize}

\end{proof}

\begin{repproposition}{thm:strong-lig}
LIG and PDA $\control$ CFG are \topeqJJ{}.
\end{repproposition}

\begin{proof}
$(\Rightarrow)$ Let $\grammar = \left(\nonterm, \alphabet, \stackalphabet, \rules, \ntvar{S}, \NTvar{S} \right)$ be a LIG in \normalform{}. Construct the controllee LD-CFG and controller PDA
\begin{align*}
\grammar_1 &= \left(\nonterm, \alphabet, \labelset, \rules_1, \ntvar{S} \right) \\
\pushdown_2 &= \left(\{q\}, \labelset, \stackalphabet, \trans_2, \left(q, \NTvar{S} \right), \left(q, \varepsilon \right) \right)
\end{align*}
The set of labels $\labelset$, the set of productions $\rules_1$ and the set of transitions $\trans_2$ are constructed as follows:
\begin{itemize}
    \item For each production $\ligprod{X}{A}{}{X}{\stackseq}{} \in \rules$, $\trans_2$ contains the transition $q, \NTvar{A} \xrightarrow{\varepsilon} q, \stackseq$.
    \item For each production $\ligprod{X}{A}{\ntvar{Y_1}[\NTvar{S}]\cdots}{Y_d}{\null}{\cdots \ntvar{Y_k}[\NTvar{S}]}$, add a fresh label $\Symvar{\ell}$ to $\labelset$, the transition $q, \NTvar{A} \xrightarrow{\Symvar{\ell}} q, \varepsilon$ to $\trans_2$ and the production $\Symvar{\ell}\colon \ntvar{X}\rightarrow \ntvar{Y}_1 \cdots \dist{\ntvar{Y}}_d \cdots \ntvar{Y}_k$ to $\rules_1$.
    \item For each production $\ligprodterm{X}{A}{a}$, add a fresh label $\Symvar{\ell}$ to $\labelset$, the transition $q, \NTvar{A} \xrightarrow{\Symvar{\ell}} q, \varepsilon$ to $\trans_2$ and the production $\Symvar{\ell}\colon \ntvar{X}\rightarrow \symvar{a}$ to $\rules_1$.
\end{itemize}

$(\Leftarrow)$ Let $\grammar_1 =\left(\nonterm, \alphabet, \labelset, \rules_1, \ntvar{S} \right)$ be a controllee LD-CFG and $\pushdown_2 = \left(\states, \labelset, \stackalphabet, \trans_2, \left(q_\init, \NTvar{S} \right), \left(q_\final, \varepsilon \right) \right)$ a controller PDA, both in \normalform{}. Construct the LIG 
\begin{align*}
\grammar' &= \left(\nonterm', \alphabet, \stackalphabet, \rules', \ntvar{S}, \NTvar{S} \right) \\
\nonterm' &= \{ \ntvar{X}^q \mid \ntvar{X}\in \nonterm, q \in \states \}
\end{align*}
and the set $\rules'$ is defined as follows:
\begin{itemize}
    \item For each transition $(p, \NTvar{A} \xrightarrow{\varepsilon} q, \stackseq) \in \trans_2$, and for each $\ntvar{X} \in \nonterm$, let $\rules'$ contain a production $\ligprod{X^p}{A}{}{X^q}{\stackseq}{}$.
    \item For each transition $(p, \NTvar{A} \xrightarrow{\Symvar{\ell}} q, \varepsilon) \in \trans_2$ and production $(\Symvar{\ell}\colon \ntvar{X}\rightarrow \ntvar{Y}_1 \cdots \dist{\ntvar{Y}}_d \cdots \ntvar{Y}_k) \in \rules_1$, let $\rules'$ contain a production $\ligprod{X^p}{A}{\ntvar{Y}_1^{q_\init} [\NTvar{S}] \cdots}{Y_d^q}{\null}{\cdots \ntvar{Y}_k^{q_\init} [\NTvar{S}]}$.
    \item For each transition $(p, \NTvar{A} \xrightarrow{\Symvar{\ell}} q_\final, \varepsilon) \in \trans_2$ and production $(\Symvar{\ell}\colon \ntvar{X}\rightarrow \symvar{a}) \in \rules_1$, let $\rules'$ contain a production $\ligprodterm{X^p}{A}{a}$. \qedhere
\end{itemize}
\end{proof}

\begin{repproposition}{thm:strong-tag}
Spinal TAG and CFG $\control$ CFG are \topeqJJ{}.
\end{repproposition}

\begin{proof}
$(\Rightarrow)$ Let $\mathbf{T} = \left(\mathcal{V}, \mathcal{C}, \alphabet, \NTvar{S}, \rules \right)$ be a spinal TAG in \normalform{}.

First, rename apart variables that allow adjunction and variables that allow substitution.

\newcommand{\whatever}{\Box}

Then construct the controllee LD-CFG and controller CFG
\begin{align*}
\grammar_1 &= (\mathcal{C} \cup \{\whatever\}, \alphabet, \labelset, \rules_1, \ntvar{S}_1)  \\
\grammar_2 &= (\mathcal{V}, \labelset, \rules_2, \NTvar{S})
\end{align*}
where $\whatever \not\in \mathcal{C}$ is a fresh constant symbol, and the set of labels $\labelset$ and the sets of rules $\rules_1$ and $\rules_2$ are constructed as follows: \david{looks kind of weird -- any way to polish it?}
\begin{itemize}

    \item For each rule of the type
        \[
        \NTvar{A} \rightarrow
        \begin{tikzpicture}[baseline=0]
        \Tree [.${\NTvar{B}_1}$ [.${\NTvar{B}_2}$ \edge[dotted,very thick]; [.${\NTvar{B}_k}\foot$ ] ] ]
        \end{tikzpicture}
        \]
        $\rules_2$ contains the production $\NTvar{A}\rightarrow \NTvar{B}_1 \NTvar{B}_2 \cdots \NTvar{B}_k$.
    \item For each rule of the type
        \[
        \NTvar{A} \rightarrow
        \begin{tikzpicture}[baseline=0]
        \Tree [.${\NTvar{B}_1}$ [.${\NTvar{B}_2}$ \edge[dotted,very thick]; [.${\NTvar{B}_k}\subst$ ] ] ]
        \end{tikzpicture}
        \]
        $L$ contains fresh labels $\ell_0$ and $\ell_k$,
        $\rules_2$ contains the production $\NTvar{S}\rightarrow \ell_0 \NTvar{B}_1 \NTvar{B}_2 \cdots \NTvar{B}_{k-1} \ell_k \NTvar{S}$, and
        $\rules_1$ contains the productions \alexandra{I am a little confused by this. I thought $\NTvar{A}$ and $\NTvar{B_k}$ are controller nonterminals \response{david} I know, everything is backward in this construction! It's because in the TAG (at least in our formulation), it's the variables that determine what can substitute where. But in the two-level grammar, the only mechanism for constraining how a non-distinguished nonterminal rewrites is the controllee nonterminal \response{david} it may have been less weird if we had defined TAG to be more like FB-TAG, where variables are of the form $X/Y$ and $X/Y$ must rewrite to a tree with root $X$ (or $X/Z$) and foot $Y$.}
        \begin{align*}
        \ell_0 \colon \NTvar{A} &\rightarrow \whatever \\
        \ell_k \colon \whatever &\rightarrow \NTvar{B}_{k}
        \end{align*}
    \item For each rule of the type
        \[
        \NTvar{A} \rightarrow \ntvar{X}\foot
        \]
        $\rules_2$ contains the production $\NTvar{A}\rightarrow \varepsilon$.
    \item For each rule of the type 
        \[
        \NTvar{A} \rightarrow \hspace{-1.5em}
        \begin{tikzpicture}[baseline=0]
        \Tree [.$\ntvar{X}$ ${\NTvar{Y}_1}\subst$ \edge[draw=none]; $\mathclap{\cdots}$ ${\NTvar{Y}_{d-1}}\subst$ $\ntvar{Z}\foot$ ${\NTvar{Y}_{d+1}}\subst$ \edge[draw=none];  $\mathclap{\cdots}$ ${\NTvar{Y}_k}\subst$ ]
        \end{tikzpicture}
        \]

        $\labelset$ contains a fresh symbol $\Symvar{\ell}$, $\rules_1$ contains the production 
        $\Symvar{\ell}\colon \whatever \rightarrow \NTvar{Y}_1 \cdots \NTvar{Y}_{d-1} \dist{\whatever} \NTvar{Y}_{d+1} \cdots \NTvar{Y}_k$ and $\rules_2$ contains the production $\NTvar{A}\rightarrow \Symvar{\ell}$.
    \item For each rule of the type 
         \[
        \NTvar{A} \rightarrow
        \begin{tikzpicture}[baseline=0]
        \Tree [.$\ntvar{X}$ $\symvar{a}$ ] 
        \end{tikzpicture}
        \]
        $\labelset$ contains a fresh symbol $\Symvar{\ell}$, $\rules_1$ contains the production 
        $\Symvar{\ell}\colon \NTvar{A} \rightarrow \symvar{a}$ and $\rules_2$ contains the production $\NTvar{S} \rightarrow \Symvar{\ell}$.
\end{itemize}

$(\Leftarrow)$ Let $\grammar_1 = (\nonterm_1, \alphabet, \labelset, \rules_1, \ntvar{S}_1)$ be an LD-CFG and $\grammar_2 = (\nonterm_2, \labelset, \rules_2, \NTvar{S}_2)$ a controller CFG. Construct the TAG 
\begin{align*}
    \mathbf{T} &= (\mathcal{V}, \nonterm, \alphabet, \NTvariii{S_2}{S_1}{}, \rules) \\
    \mathcal{V} &= \{ \NTvariii{A}{X}{Y} \mid \ntvar{X}, \ntvar{Y} \in \nonterm, \NTvar{A} \in \nonterm' \} \\
    &\quad \cup \{ \NTvariii{A}{X}{} \mid \ntvar{X} \in \nonterm, \NTvar{A} \in \nonterm' \}
\end{align*}
and the set $\rules$ is constructed as follows:
\begin{itemize}
    \item For each production $\NTvar{A} \rightarrow \NTvar{B}_1 \cdots \NTvar{B}_k \in \rules_2$, where $k>0$, and for each $\ntvar{X_0}, \ntvar{X_1}, \ldots, \ntvar{X_k} \in \nonterm$, let $\rules$ contain the rules
    \begin{align*} \hspace{-1em}
    \NTvariii{A}{X_0}{X_k} &\rightarrow
    \begin{tikzpicture}[baseline=0]
    \Tree [.$\NTvariii{B_1}{X_0}{X_1}$ [.$\treevdots$ [.$\mathclap{\NTvariii{B_k}{X_{k-1}}{X_k}\foot}$ ] ] ] 
    \end{tikzpicture}
    &
    \NTvariii{A}{X_0}{} &\rightarrow
    \begin{tikzpicture}[baseline=0]
    \Tree [.$\NTvariii{B_1}{X_0}{X_1}$ [.$\treevdots$ [.$\mathclap{\NTvariii{B_k}{X_{k-1}}{}\subst}$ ] ] ] 
    \end{tikzpicture}
    \end{align*}
    \item For each production $\NTvar{A} \rightarrow \varepsilon \in \rules_2$, let $\rules$ contain the rules
    \begin{align*}
    \NTvariii{A}{X}{X} &\rightarrow \ntvar{X}\foot & \NTvariii{A}{X}{} &\rightarrow \Tree [.$\ntvar{X}$ $\varepsilon$ ]
    \end{align*}
    \item For each pair of productions $\NTvar{A}\rightarrow \Symvar{\ell}\in \rules_2$ and $\Symvar{\ell}\colon \ntvar{X}\rightarrow \symvar{a} \in \rules_1$, let $\rules$ contain the rule 
    \[
    \NTvariii{A}{X}{} \rightarrow
    \begin{tikzpicture}[baseline=0]
    \Tree [.$\ntvar{X}$ [.$\symvar{a}$ ] ] 
    \end{tikzpicture}
    \]
    \item For each pair of productions $\NTvar{A}\rightarrow \Symvar{\ell}\in \rules_2$ and $\Symvar{\ell}\colon \ntvar{X}\rightarrow \ntvar{Y_1} \cdots \ntvar{Y_{d-1}} \dist{\ntvar{Z}} \ntvar{Y_{d+1}} \cdots \ntvar{Y_k} \in \rules_1$, let $\rules$ contain the rule 
    {\small
    \[\hspace{-1em}
    \NTvariii{A}{X}{Z}\rightarrow \hspace*{-3em}
    \begin{tikzpicture}[baseline=0,sibling distance=0]
    \Tree [.{$\ntvar{X}$} {${\NTvariii{S_2}{Y_1}{}}\subst$}  \edge[draw=none]; {$\treecdots$} {${\NTvariii{S_2}{Y_{d-1}}{}}\subst$} {$\ntvar{Z}\foot$} {${\NTvariii{S_2}{Y_{d+1}}{}}\subst$} \edge[draw=none]; {$\treecdots$} {${{\NTvariii{S_2}{Y_k}{}}\subst}$} ]
\end{tikzpicture}
    \]}

\end{itemize}
\end{proof}

\begin{repproposition}{thm:strong-paa}
Spinal PAA and CFG $\control$ PDA are \topeqJJ{}.
\end{repproposition}

\begin{proof}
$(\Rightarrow)$ Let $\mathbf{M} = (\alphabet, \mathcal{V}, \mathcal{C}, \trans, \NTvar{S})$ be a PAA in \normalform{}. 

First, rename apart variables that allow adjunction and variables that allow substitution.

Then construct the LD-PDA controllee and CFG controller
\begin{align*}
\pushdown_1 &= (\{q\}, \alphabet, \mathcal{C} \cup \{ \Box \}, \labelset, \trans_1, \ntvar{S}_1) \\
\grammar_2 &= (\mathcal{V}, \labelset, \rules_2, \NTvar{S})
\end{align*}
where $\Box \notin \mathcal{C}$ is a fresh constant symbol, and the sets $\labelset$, $\trans_1$ and $\rules_2$ are constructed as follows:
\begin{itemize}
    \item For every transition in $\trans$ of the form 
    \begin{equation*}
        \tikzset{paa picture}
        \begin{tikzpicture} \node(n1){$\NTvar{A}$}; \fnode{n1} \end{tikzpicture}
        \xrightarrow{\varepsilon} 
        \begin{tikzpicture}
        \node(n1){$\NTvar{Y}_1$};
        \rnode{n1}{n2}{${}\cdots{}$}
        \rnode{n2}{n3}{$\NTvar{Y}_{d-1}$}
        \rnode{n3}{n4}{$\,\ntvar{Z}\,$}
        \fnode{n4}
        \rnode{n4}{n5}{$\NTvar{Y}_{d+1}$}
        \rnode{n5}{n6}{${}\cdots{}$}
        \rnode{n6}{n7}{$\NTvar{Y}_k$}
        \end{tikzpicture}
    \end{equation*}
    $\labelset$ contains a fresh label $\Symvar{\ell}$, $\rules_2$ contains the production $\NTvar{A}\rightarrow \Symvar{\ell}$ and $\trans_1$ contains the transition $\Symvar{\ell}\colon q, \Box \xrightarrow{\varepsilon} q, \NTvar{Y}_1 \cdots \NTvar{Y}_{d-1} \dist{\Box} \NTvar{Y}_{d+1} \cdots \NTvar{Y}_k$.
    \item For every transition in $\trans$ of the form 
    \[
    \NTvar{A} \xrightarrow{\symvar{a}} 
    \varepsilon \]
    $\labelset$ contains a fresh label $\Symvar{\ell}$, $\rules_2$ contains the production $\NTvar{S}\rightarrow \Symvar{\ell}$ and $\trans_1$ contains the transition $\Symvar{\ell}\colon q, \NTvar{A}\xrightarrow{\symvar{a}} q, \varepsilon$.
    \item For every transition in $\trans$ of the form
    \[\tikzset{paa picture}
    \begin{tikzpicture}
    \node(n1){$\NTvar{A}$};
    \fnode{n1}
    \end{tikzpicture} 
    \xrightarrow{\varepsilon} 
    \begin{tikzpicture}
    \node(n1){$\NTvar{B_1}$};
    \dnode{n1}{n2}{$\NTvar{B_2}$}
    \dotsnode{n2}{n3}{$\NTvar{B_k}$}
    \fnode{n3}
    \end{tikzpicture}
    \]
    $\rules_2$ contains the production $\NTvar{A}\rightarrow \NTvar{B_1}\cdots \NTvar{B_k}$.
    \item For every transition in $\trans$ of the form
    \[\tikzset{paa picture}
    \begin{tikzpicture}
    \node(n1){$\NTvar{A}$};
    \end{tikzpicture} 
    \xrightarrow{\varepsilon} 
    \begin{tikzpicture}
    \node(n1){$\NTvar{B_1}$};
    \dnode{n1}{n2}{$\NTvar{B_2}$}
    \dotsnode{n2}{n3}{$\NTvar{B_k}$}
    \end{tikzpicture}
    \]
    $\labelset$ contains fresh labels $\ell_0$ and $\ell_k$, 
    $\rules_2$ contains the production $\NTvar{S}\rightarrow \ell_0 \NTvar{B_1}\cdots \NTvar{B_{k-1}} \ell_k \NTvar{S}$ and $\trans_1$ contains the transitions
    \begin{align*}
        \ell_0 \colon q, \NTvar{A} &\xrightarrow{\varepsilon} q, \Box \\
        \ell_k \colon q, \Box &\xrightarrow{\varepsilon} q, \NTvar{B_k}
    \end{align*}
    \item For every transition in $\trans$ of the form 
    \[\tikzset{paa picture}
    \begin{tikzpicture}\node(n1){$\NTvar{A}$}; \fnode{n1}\end{tikzpicture} \xrightarrow{\varepsilon} 
    \begin{tikzpicture}\node(n1){$\ntvar{X}$}; \fnode{n1}\end{tikzpicture}
    \]
    $\rules_2$ contains the production $\NTvar{A}\rightarrow \varepsilon$.
\end{itemize}

$(\Leftarrow)$
We are given $\grammar_2 \control \pushdown_1$, where $\grammar_2$ is a CFG and $\pushdown_1$ is an LD-PDA. Without loss of generality, we may assume by \cref{thm:onestatepda} that $\pushdown_1$ has only one state, so
\begin{align*}
\pushdown_1 &= (\{q \}, \alphabet, \stackalphabet, \labelset, \trans_1, (q_\init, \ntvar{S}_1), (q_\final, \varepsilon)) \\
\grammar_2 &= (\nonterm, \labelset, \rules_2, \NTvar{S}_2)
\end{align*}
We further assume that both are in normal form.
Then we construct the PAA 
\begin{align*}
\mathbf{M} &= (\alphabet, \mathcal{V}, \stackalphabet, \trans', \NTvariii{S_2}{S_1}{}) \\
\mathcal{V} &= \{ \NTvariii{A}{X}{Y} \mid \NTvar{A} \in \nonterm, \ntvar{X},\ntvar{Y}\in \stackalphabet \} \\
&\quad \cup  \{ \NTvariii{A}{X}{} \mid \NTvar{A}\in \nonterm, \ntvar{X} \in \stackalphabet \}
\end{align*}
where $\trans'$ is constructed as follows:
\begin{itemize}
    \item For each production $\NTvar{A}\rightarrow \NTvar{B}_1 \cdots \NTvar{B}_k \in \rules_2$, where $k>0$, 
    and for all $\ntvar{X}_0, \ntvar{X}_1, \ldots, \ntvar{X}_k \in \stackalphabet$,
    let $\trans'$ contain a transition  
    \[\tikzset{paa picture} \small
    \begin{tikzpicture}
    \node(n1){$\NTvariii{A}{{X_0}}{{X_k}}$};
    \fnode{n1}
    \end{tikzpicture} 
    \xrightarrow{\varepsilon}
    \begin{tikzpicture}
    \node(n1){$\NTvariii{B_1}{{X_0}}{{X_1}}$};
    \dnode{n1}{n2}{$\NTvariii{B_2}{{X_1}}{{X_2}}$}
    \dotsnode{n2}{n3}{$\NTvariii{B_k}{{X_{k-1}}}{{X_k}}$}
    \fnode{n3}
    \end{tikzpicture}
    \]
    \item For each production $\NTvar{A}\rightarrow \varepsilon \in \rules_2$ and for all $\ntvar{X} \in \stackalphabet$, let $\trans'$ contain a transition 
    \[ \tikzset{paa picture}
    \begin{tikzpicture}\node(n1){$\NTvariii{A}{X}{X}$}; \fnode{n1}\end{tikzpicture} \xrightarrow{\varepsilon} \begin{tikzpicture}\node(n1){$\ntvar{X}$}; \fnode{n1}\end{tikzpicture}
    \]
    \item For all pairs of rules $\NTvar{A}\rightarrow \Symvar{\ell}\in\rules_2$ and $\Symvar{\ell}\colon q,\ntvar{X} \xrightarrow{\symvar{a}} q, \varepsilon \in \trans_1$, let $\trans'$ contain the transition $ \NTvariii{A}{X}{} \xrightarrow{\symvar{a}} \varepsilon$. 
    \item For all pairs of rules $\NTvar{A}\rightarrow \Symvar{\ell}\in \rules_2$ and 
    $\Symvar{\ell} \colon q, \ntvar{X} \xrightarrow{\varepsilon} q, \ntvar{Y}_1 \cdots \ntvar{Y}_{d-1} \dist{\ntvar{Z}} \ntvar{Y}_{d+1} \cdots \ntvar{Y}_k \in \trans_1$, let $\trans'$ contain a transition 
    \[ \tikzset{paa picture} \footnotesize
    \begin{tikzpicture} \node(n1){$\NTvariii{A}{X}{Z}$}; \fnode{n1} \end{tikzpicture}
    \xrightarrow{\varepsilon} \begin{tikzpicture}
    \node(n1){$\NTvariii{S_2}{Y_1}{}$};
    \rnode{n1}{n2}{${}\cdots{}$};
    \rnode{n2}{n3}{$\NTvariii{S_2}{Y_{d-1}}{}$};
    \rnode{n3}{n4}{$\,\ntvar{Z}\,$});
    \fnode{n4};
    \rnode{n4}{n5}{$\NTvariii{S_2}{Y_{d+1}}{}$};
    \rnode{n5}{n6}{${}\cdots{}$};
    \rnode{n6}{n7}{$\NTvariii{S_2}{Y_k}{}$};
    \end{tikzpicture}
    \]
\end{itemize}
\end{proof}

\end{document}